\documentclass[a4paper, 11pt]{article}
\pdfoutput=1
\usepackage{jheppub}
\usepackage{graphicx}

\usepackage{wrapfig,float,slashed}
\usepackage{amsmath,dsfont}
\usepackage{amssymb,epsfig,graphicx}
\usepackage[table]{xcolor}
\usepackage{epstopdf}
\usepackage{tabu}
\usepackage[normalem]{ulem}
\usepackage{ragged2e}
\usepackage{lipsum}
\usepackage{siunitx}
\usepackage[font=small,skip=1pt]{caption}
\allowdisplaybreaks
\usepackage{pifont}
\newcommand{\cmark}{\ding{51}}
\newcommand{\xmark}{\ding{55}}
\usepackage[hang,flushmargin,multiple]{footmisc}

\usepackage{tikz}
\usetikzlibrary{arrows,positioning,shapes,mindmap}
\usetikzlibrary{decorations.pathmorphing}
\usetikzlibrary{decorations.markings}

\newcommand{\nc}{\newcommand}
\nc{\non}{\nonumber}
\nc{\hc}{\hbox {H.c.}}
\nc{\noi}{\noindent}
\nc{\barx}{\bar{x}}
\nc{\pbarn}{\;\hbox {pb}}
\nc{\fbarn}{\;\hbox {fb}}

\nc{\hsp}{\hspace{0.5cm}}
\nc{\lsp}{\hspace{1cm}}
\nc{\Lsp}{\hspace{2cm}}
\nc{\LLsp}{\lsp\lsp}
\nc{\lra}{\longrightarrow}
\nc{\p}{\prime}
\nc{\sgn}{\text{sgn}}
\nc{\arccot}{\text{arccot}}
\nc{\ph}{\varphi}
\nc{\co}{{\cal O}}
\nc{\beq}{\begin{equation}}  \nc{\eeq}{\end{equation}}
\nc{\bea}{\begin{eqnarray}}  \nc{\eea}{\end{eqnarray}}
\nc{\baa}{\begin{array}}     \nc{\eaa}{\end{array}}
\nc{\bit}{\begin{itemize}}   \nc{\eit}{\end{itemize}}
\nc{\ben}{\begin{enumerate}} \nc{\een}{\end{enumerate}}
\nc{\bce}{\begin{center}}    \nc{\ece}{\end{center}}
\nc{\bpm}{\begin{pmatrix}}   \nc{\epm}{\end{pmatrix}}
\nc{\bvt}{\begin{verbatim}}  \nc{\evt}{\end{verbatim}}

\def\lsim{\mathrel{\raise.3ex\hbox{$<$\kern-.75em\lower1ex\hbox{$\sim$}}}}
\def\gsim{\mathrel{\raise.3ex\hbox{$>$\kern-.75em\lower1ex\hbox{$\sim$}}}}

\def\udots{\mathinner{\mkern1mu\raise1pt\vbox{\kern7pt\hbox{.}}\mkern2mu\raise4pt\hbox{.}\mkern2mu\raise7pt\hbox{.}\mkern1mu}}

\def\gev{\;\hbox{GeV}}
\def\tev{\;\hbox{TeV}}
\def\eq{\hbox{Eq.~}}
\def\mpl{M_{\text{Pl}}}
\definecolor{agray}{rgb}{0.95, 0.95, 0.99}

\newcommand\fverb{\setbox\fverbbox=\hbox\bgroup\verb}
\newcommand\fverbdo{\egroup\medskip\noindent%
			\fbox{\unhbox\fverbbox}\ }
\newcommand\fverbit{\egroup\item[\fbox{\unhbox\fverbbox}]}
\newbox\fverbbox

\makeatletter
\renewcommand{\boxed}[1]{\textcolor{black}{%
\tikz[baseline={([yshift=-0ex]current bounding box.center)}] \node [rectangle, minimum width=0ex,draw] {\normalcolor\m@th$\displaystyle#1$};}}
 \makeatother

\title{Higgs Dark Matter from a Warped Extra Dimension\\
-- {\it the truncated-inert-doublet model}}
\author[a,b,c]{Aqeel Ahmed,}
\author[a]{Bohdan Grzadkowski,}
\author[b]{John F. Gunion}
\author[b]{and Yun Jiang}
\affiliation[a]{Faculty of Physics,
University of Warsaw,\\
Pasteura 5, 02-093 Warsaw, Poland}
\affiliation[b]{Department of Physics, University of California,\\
Davis, CA 95616, U.S.A.}
\affiliation[c]{National Centre for Physics, Quaid-i-Azam University Campus,\\
Shahdra Valley Road, Islamabad 45320, Pakistan}
\emailAdd{aqeel.ahmed@fuw.edu.pl}
\emailAdd{bohdan.grzadkowski@fuw.edu.pl}
\emailAdd{jfgunion@ucdavis.edu}
\emailAdd{yunjiang@ucdavis.edu}

\abstract{
We construct a 5D $\mathbb{Z}_2$-symmetric model with three D3-branes: two IR ones with negative tension
located at the ends of an extra-dimensional interval and a UV-brane with positive tension
placed in the middle of the interval -- IR-UV-IR model. The background solutions for this geometric setup are found without and with taking into account the backreaction of the matter fields. A 5D $SU(2)$ Higgs doublet is employed as the Goldberger-Wise stabilizing field in this geometry and solutions of the 5D coupled scalar-gravity equations are found by using the superpotential method. Within this setup we
investigate the low-energy (zero-mode) effective theory for the bulk Standard Model (SM) bosonic sector. The $\mathbb{Z}_2$-even zero-modes correspond to known standard degrees of freedom, whereas the
$\mathbb{Z}_2$-odd zero modes might serve as a dark sector.
The effective low-energy scalar sector contains a scalar which mimics the SM Higgs boson and a second stable scalar particle (dark-Higgs) is a dark matter candidate; the latter  is a component of the zero-mode of the $\mathbb{Z}_2$-odd Higgs doublet. The model that results from  the $\mathbb{Z}_2$-symmetric background geometry resembles the Inert Two Higgs Doublet Model.
The effective theory turns out to have an extra residual $SU(2)\times U(1)$ global symmetry that is reminiscent of an underlying 5D gauge transformation for the odd degrees of freedom.
At tree level the SM Higgs and the dark-Higgs  have the same mass; however, when leading radiative corrections are taken into account the dark-Higgs turns out to be heavier than the SM Higgs. Implications for dark matter are discussed; it is found that the dark-Higgs can provide only a small fraction of the observed dark matter abundance.
}

\arxivnumber{1504.03706}

\begin{document}

\maketitle
\flushbottom

\tikzstyle{every picture}+=[remember picture]
\pgfdeclaredecoration{complete sines}{initial}
{
    \state{initial}[
        width=+0pt,
        next state=sine,
        persistent precomputation={\pgfmathsetmacro\matchinglength{
            \pgfdecoratedinputsegmentlength / int(\pgfdecoratedinputsegmentlength/\pgfdecorationsegmentlength)}
            \setlength{\pgfdecorationsegmentlength}{\matchinglength pt}
        }] {}
    \state{sine}[width=\pgfdecorationsegmentlength]{
        \pgfpathsine{\pgfpoint{0.25\pgfdecorationsegmentlength}{0.5\pgfdecorationsegmentamplitude}}
        \pgfpathcosine{\pgfpoint{0.25\pgfdecorationsegmentlength}{-0.5\pgfdecorationsegmentamplitude}}
        \pgfpathsine{\pgfpoint{0.25\pgfdecorationsegmentlength}{-0.5\pgfdecorationsegmentamplitude}}
        \pgfpathcosine{\pgfpoint{0.25\pgfdecorationsegmentlength}{0.5\pgfdecorationsegmentamplitude}}
}
    \state{final}{}
}
\tikzset{
fermion/.style={thick,draw=black, line cap=round, postaction={decorate},
    decoration={markings,mark=at position 0.6 with {\arrow[black]{latex}}}},
photon/.style={thick, line cap=round,decorate, draw=black,
    decoration={complete sines,amplitude=4pt, segment length=6pt}},
boson/.style={thick, line cap=round,decorate, draw=black,
    decoration={complete sines,amplitude=4pt,segment length=8pt}},
gluon/.style={thick,line cap=round, decorate, draw=black,
    decoration={coil,aspect=1,amplitude=3pt, segment length=8pt}},
scalar/.style={dashed, thick,line cap=round, decorate, draw=black},
ghost/.style={dotted, thick,line cap=round, decorate, draw=black},
->-/.style={decoration={
  markings,
  mark=at position 0.6 with {\arrow{>}}},postaction={decorate}}
 }

\makeatletter
\tikzset{
    position/.style args={#1 degrees from #2}{
        at=(#2.#1), anchor=#1+180, shift=(#1:\tikz@node@distance)
    }
}
\makeatother

\section{Introduction}
\label{Introduction}

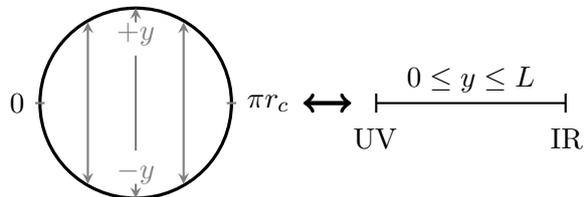
\begin{wrapfigure}{r}{0.5\textwidth}
\centering
\begin{tikzpicture}[very thick,scale=0.63]
\draw[black] (0,0)circle (2cm);
\draw[gray, thick]    (-2.1,0) node[left,black]{$0$} -- (-1.9,0)
                (2.1,0)node[right,black]{$\pi r_c$} -- (1.9,0);
\draw[<-,>=stealth,thick,gray] (0,2) -- (0,1.7);
\draw[<-,>=stealth,thick,gray] (0,-2) -- (0,-1.7);
\draw[<->,>=stealth,thick,gray] (1,1.7) -- (1,-1.7);
\draw[<->,>=stealth,thick,gray] (-1,1.7) -- (-1,-1.7);
\draw[thick, gray] (0,1)node[above]{$+y$} -- (0,-1)node[below]{$-y$};
\draw[<->, ultra thick] (3.5,0) --(4.5,0);
\draw[|-|,thick] (5,0)node[below=2mm]{UV} --  node[above]{$0\leq y\leq L$} (9,0)node[below=2mm]{IR};
\end{tikzpicture}
\caption{Cartoon of RS1 geometry.}
\label{fig_rs_geometry}
\end{wrapfigure}
The seminal work of Randall and Sundrum (RS) \cite{Randall:1999ee} provides an elegant solution to {\it the hierarchy problem}. Their proposal involves an extra-dimension with a non-trivial warp factor due to the assumed anti-de Sitter (AdS) geometry along the extra-dimension. In their model an AdS geometry on an $S_1/\mathbb{Z}_2$ orbifold is considered which is equivalent to a line-element $0\leq y\leq L$, where $y$ is the coordinate of the fifth-dimension and $L=\pi r_c$,
with $r_c$ being the radius of the circle in the fifth-dimension. Moreover, their model involves two D3-branes localized on the fixed points of the orbifold, a ``UV-brane'' at $y=0$ and  an ``IR-brane'' at $y=L$ (our nomenclature will become clear below), see Fig. \ref{fig_rs_geometry}.
The solution for the RS geometry is \cite{Randall:1999ee,Randall:1999vf},
\beq
ds^2=e^{2A(y)}\eta_{\mu\nu}dx^\mu dx^\nu+dy^2, \lsp \text{with}\lsp A(y)=-k|y|, \label{metric}
\eeq
where $k$ is the inverse of the AdS radius. In the original RS1 model \cite{Randall:1999ee} it was assumed that the Standard Model (SM) is localized on the IR-brane, whereas gravity is localized on the UV-brane and  propagates through the bulk to the IR-brane. They famously showed that if the 5D fundamental theory involves only one mass scale
$M_\ast$ -- the Planck mass in 5D -- then, due to the presence of non-trivial warping along the extra-dimension, the effective mass scale on the IR-brane is rescaled to $m_{KK}\sim ke^{-kL}\sim\co(\tev)$ and hence ameliorates the hierarchy problem for mild values of $kL\sim\co(35)$.

Soon after the RS proposal, many important improvements to the model were considered. First, a stabilization mechanism for the RS1 setup was proposed by Goldberger and Wise \cite{Goldberger:1999uk}; it employs a real scalar field in the bulk of AdS geometry with localized potentials on both of the branes, see also \cite{DeWolfe:1999cp}. A second interesting observation, which could potentially solve the fermion mass hierarchy problem within the SM, was made by many groups \cite{Davoudiasl:1999tf,Grossman:1999ra,Chang:1999nh,Gherghetta:2000qt,Huber:2000ie}. The core idea of these works was to allow all the SM fields to propagate in the RS1 bulk, except the Higgs field which was kept localized on the IR-brane. In this way, the zero-modes of these bulk fields correspond to the SM fields and the overlap of $y$-dependent profiles of fermionic fields with the Higgs field could generate the required fermion mass hierarchy. To suppress the electroweak (EW) precision observables, the symmetry of the gauge group was enhanced by introducing custodial symmetry in Ref. \cite{Agashe:2003zs}. The common lore, in the RS1 model and its extensions, was to keep the Higgs field localized on the IR-brane in order to solve the hierarchy problem. The first attempt to consider the Higgs field in the bulk of RS1 was made by Luty and Okui \cite{Luty:2004ye}. They employed AdS/CFT duality~\footnote{For the phenomenological applications of AdS/CFT with RS1 geometry, see for example \cite{ArkaniHamed:2000ds,Rattazzi:2000hs}.} to argue that a {\it bulk Higgs} scenario can address the hierarchy problem by making the Higgs mass operator marginal in the dual CFT.

A study of electroweak symmetry breaking (EWSB) within the bulk Higgs scenario was first performed in the RS1 setup by Davoudiasl et al. \cite{Davoudiasl:2005uu}; they showed that the zero-mode of the bulk Higgs is tachyonic and hence could lead to a vacuum expectation value (vev) at the TeV scale. Recently there have been many studies where a bulk Higgs scenario is considered from different perspectives ---  see for example: a study with custodial symmetry in the Higgs sector\cite{Cacciapaglia:2006mz}; models with a soft wall setup \cite{Falkowski:2008fz}; bulk Higgs mediated FCNC's \cite{Azatov:2009na}; suppression of EW precision observables by modifying the warped metric near the IR-brane \cite{Cabrer:2010si,Cabrer:2011fb,Cabrer:2011vu}; and, a bulk Higgs as the modulus stabilization field (Higgs--radion unification) \cite{Geller:2013cfa}. Different phenomenological aspects after the Higgs discovery were explored in \cite{Archer:2012qa,Frank:2013un,Malm:2013jia,Cox:2013rva,Archer:2014jca,Dillon:2014zea,Agashe:2014jca,Iyer:2015ywa}. These phenomenological studies show that the RS1 model with bulk SM fields and its descendants with modified geometry (RS-like warped geometries in general) are consistent with the current experimental bounds and EW precision data.

A separate category of generalization of the RS models is based on the assumption that the singular branes are replaced
with thick branes which are smooth field configurations of the bulk scalar field, see e.g. \cite{DeWolfe:1999cp} and
\cite{Ahmed:2012nh,Ahmed:2013mea}.

As we discussed above, RS-like warped geometries, being consistent with the experimental data, offer an attractive solution to many of the fundamental puzzles of the SM, mostly through geometric means. In the same spirit, one can ask if RS-like warped extra-dimensions can shed some light on another outstanding puzzle of SM, the lack of a candidate for dark matter (DM) which constitutes 83\% of the observed matter in the universe \cite{Planck:2015xua}. It appears that unlike (flat) universal extra-dimensions (UED), where the KK-modes of the bulk fields can be even and odd under KK-parity (implying that the lowest KK-odd particle (LKP) could be a natural dark matter candidate \cite{Servant:2002aq,Cheng:2002ej}), RS1-like models (involving two branes and warped bulk) are unable to offer an analogue of KK-parity. The reason lies in the fact that the RS1 geometry is just a single slice of AdS space and, since warped, cannot be symmetric around any point along the extra-dimension and hence does not allow a {\it KK-parity}. As a result it cannot accommodate a realistic dark matter candidate. To cure this problem in the warped geometries, usually extra discrete symmetries are introduced such that the SM fields are even while the DM is odd under such discrete symmetries in order to make it stable \cite{Agashe:2004ci,Panico:2008bx,Ponton:2008zv,Vecchi:2013xra}. Another way to mend this problem in warped geometries is to introduce an additional {\it hidden sector} with some local gauge symmetries such that only DM is charged under the hidden sector gauge symmetries and it couples to the SM very weakly \cite{Gherghetta:2010cq,vonHarling:2012sz}, (see also \cite{Frey:2009qb}).

\begin{figure}
\centerline{
\begin{tikzpicture}[very thick,rounded corners=0.5pt,line cap=round,scale=0.6]
\shadedraw[top color=blue!50,bottom color=blue!10,yslant=0.1]
(0,0) -- (2,2) -- node[above=-13pt,rotate=-90]{\small UV-brane} (2,7) -- (0,5) --  cycle;
\shadedraw[bottom color=orange!75,top color=orange!10,yslant=0.1](8,0.5) -- (9,1.5) -- node[above,rotate=-90]{\small IR-brane} (9,4) -- (8,3) -- cycle;
\shadedraw[bottom color=orange!75,top color=orange!10,yslant=0.1](-7,2.1) -- (-6,3.1) -- node[above=-1pt,rotate=-90]{\small IR-brane} (-6,5.6) -- (-7,4.6) -- cycle;
\draw[thick,black,yslant=0.1,opacity=0.5](0,5) to [out=-30,in=170]node[above right,blue,opacity=1]{\small $e^{-2k|y|}$} (8,3)
                                (2,7) to [out=-50,in=170] (9,4)
                                (2,2) to [out=8,in=180] (9,1.5)
                                (0,0) to [out=20,in=175] (8,0.5);
\draw[thick,black,yslant=0.1,opacity=0.5](0,5) to [out=200,in=0]node[above,blue,opacity=1]{\small $e^{-2k|y|}$} (-7,4.6)
                                (2,7) to [out=210,in=5] (-6,5.6)
                                (2,2) to [out=165,in=-5] (-6,3.1)
                                (0,0) to [out=145,in=-5] (-7,2.1);
\draw (0,0) node[below,blue]{\small $y=0$}
 (8,0.5)node[above,blue]{\small $y=L$}
 (-7,0.5)node[above,blue]{\small $y=-L$}
 (5,2.5)node[above,black]{\small $\Lambda_B$}
 (-3,2.5)node[above,black]{\small $\Lambda_B$};
\draw[->,>=stealth,thick,yslant=0.1] (1,1) -- (1,2.1) node[above]{\small $x^\mu$};
\draw[->,>=stealth,thick,yslant=0.1](1,1)-- (2.1,1) node[right]{\small $y$};
\end{tikzpicture}
}
\caption{The geometric configuration for IR-UV-IR setup, the parameters are defined in Sec. \ref{IR-UV-IR background solution}.}
\label{IR-UV-IR}
\end{figure}
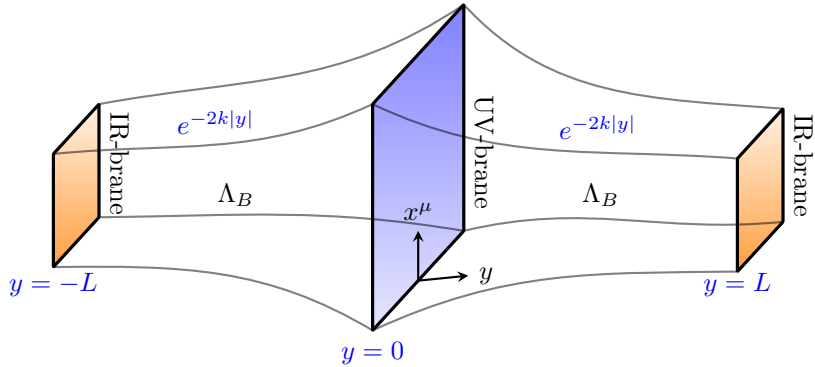

An alternative to introducing additional symmetries, is to extend the RS1-like warped geometry in such a way that the whole geometric setup becomes symmetric around a fixed point in the bulk. Two $\mathbb{Z}_2$ symmetric warped configurations are possible. In the first,  two identical AdS patches are symmetrically  glued together at a UV fixed point,  while in the second  two identical AdS pathes are symmetrically glued together at an IR fixed point. The geometric configuration when the
two AdS copies are glued together at the UV fixed point will be referred as ``IR-UV-IR geometry'', whereas the
geometry corresponding to the setup when two AdS copies are glued at the IR fixed point is called ``UV-IR-UV geometry''. We will only consider the IR-UV-IR geometric setup --- it is straight forward to extend our analysis to the UV-IR-UV geometries. (A common pathology associated with this latter type of geometry is the appearance of ghosts.) We consider an interval $y\in[-L, L]$ in the extra-dimension, where on each end of the interval $y=\pm L$ there is a D3 brane with negative tension
(in Sec. \ref{IR-UV-IR background solution} it will be clear why we need negative tension branes) and at the center
of the interval,  $y=0$, we place a positive tension brane where we assume that gravity is localized~\footnote{One can {\it smooth} the singular branes in our setup by appropriate scalar field configurations --- for smooth brane modeling see for example \cite{Ahmed:2012nh} and references therein.}. We call the boundary
branes ``IR-branes'' and the brane at $y=0$ we term the ``UV-brane''. The IR-UV-IR geometry and a pictorial description of such a geometric setup is shown in Fig. \ref{IR-UV-IR}. Since the brane tensions of the two IR-branes are the same, this geometry is $\mathbb{Z}_2$ symmetric. We are aware of only two earlier attempts to construct a similar setup. The first \cite{Agashe:2007jb} treated the lowest odd KK gauge mode as the DM candidate. The second employed a kink-like UV thick brane  \cite{Medina:2010mu} and the corresponding dark-matter was the first odd KK-radion \cite{Medina:2011qc}.

In this work, we place all the SM fields, including the Higgs doublet, in the bulk of the IR-UV-IR geometry. We calculate the background solutions for our geometric setup without and with taking into account the backreaction of matter fields. Since only 5D Higgs doublet, present in the bulk as well on the branes, acquires $y$-dependent vev, therefore we solve the full 5D scalar-gravity coupled set of Einstein equations to get solutions which address the gauge hierarchy problem. Here 5D $SU(2)$ Higgs doublet plays the role of the Goldberger-Wise stabilization field and the values of Higgs vevs at the UV- and IR-brane fix the distance between the branes.  We find that for a weak backreaction, the UV-brane Higgs vev has to be much smaller than that of the IR-brane. Moreover we show that in order to have 4D cosmological constant zero at the IR branes one needs precisely one fine-tuning, similar to RS1 \cite{Goldberger:1999uk,DeWolfe:1999cp}.

The geometric $\mathbb{Z}_2$ parity ($y\to-y$ symmetry) leads to ``warped KK-parity'',    i.e.  there are towers of even and
odd KK-modes corresponding to each bulk field. In the weak backreaction scenario we focus on EWSB induced by the bulk Higgs doublet and low energy aspects of the 4D effective theory for the even and odd zero-modes assuming the KK-mass scale is high enough $\sim\co(\text{few})\tev$. In the effective theory the even and odd Higgs doublets mimic a two-Higgs-doublet model (2HDM) scenario -- the truncated inert-doublet model -- with the odd doublet similar to the inert doublet but without corresponding pseudoscalar and charged scalars. All the parameters of this {\it truncated} 2HDM are determined by the fundamental 5D parameters of the theory and the choice of boundary conditions (b.c.) for the fields at $y=\pm L$. (Note that the boundary or ``jump'' conditions at $y=0$ follow from the bulk equations of motion in the case of even modes, whereas odd modes are required to be zero by symmetry.) There are many possible alternative choices for the b.c. at $\pm L$. We allow the $y$-derivative of a field to have an arbitrary value at $\pm L$ as opposed to requiring that the field value itself be zero, i.e. we employ Neumann or mixed b.c.  rather than Dirichlet b.c.  at $\pm L$.  Only the former yields a non-trivial theory allowing spontaneous symmetry breaking (SSB), whereas the latter leads to an explicit symmetry breaking scenario in which there are no Goldstone modes and the gauge bosons do not acquire mass. With these choices, the symmetric setup yields an odd Higgs zero-mode that is a natural candidate for dark matter. We compute the one-loop quadratic (in cutoff) corrections to the two scalar zero modes within the effective theory and discuss their mass splitting. The dark matter candidate is a WIMP --- we calculate its relic abundance in the cold dark matter paradigm.

The paper is composed as follows. In Sec. \ref{KK-parity from warped geometry}, we setup the IR-UV-IR geometric configuration and provide background solutions without and with backreaction due to the presence of matter contents in the bulk and on the branes. We also discuss the manifestation of KK-parity due the $\mathbb{Z}_2$ geometric setup. Section \ref{SM EWSB due to bulk Higgs doublet} contains the main part of our work. There, we focus on EWSB for the SM gauge sector due to the bulk Higgs doublet in our ${\mathbb Z}_2$ symmetric geometry and obtain a low-energy 4D effective theory containing all the SM fields plus a real scalar -- a dark matter candidate -- which is odd under the discrete $\mathbb{Z}_2$ symmetry. In the subsequent two subsections of Sec. \ref{SM EWSB due to bulk Higgs doublet}, we consider the quantum corrections to the scalar masses below the KK-scale $\sim\co(\text{few})\tev$ and explore the possible implications of the dark-matter candidate by calculating its relic abundance. We summarize and give our conclusions in Sec. \ref{Summary}. We supplement the main text with an Abelian Higgs mechanism, with a complex scalar field and a gauge field, in our background geometry in Appendix \ref{SSB in the IR-UV-IR model: the Abelian Higgs mechanism}. In the Abelian case we lay down the foundation for SSB due to bulk Higgs, which is useful in the main text for the case of EWSB of the SM. Two apparently different approaches are considered to study SSB in the Abelian case: ({\it i}) SSB by vacuum expectation values of the KK modes; and, ({\it ii}) SSB via a vacuum expectation value of the 5D Higgs field. Low energy (zero-mode) 4D effective theories are obtained within the two approaches and we find that the effective theories are identical up to corrections of order $\co\big(m_0^2/m_{KK}^2\big)$, where $m_0$ and $m_{KK}$ are the zero-mode mass and KK-mass scale, respectively.

\section{A ${\mathbb Z}_2$ symmetric warped extra-dimension and KK-parity}
\label{KK-parity from warped geometry}
In this section we provide the background solution for the $\mathbb{Z}_2$ symmetric background (IR-UV-IR) geometry without and with backreaction due to the 5D matter fields and show how KK-parity is manifested within this warped geometric setup.

\subsection{The IR-UV-IR model: without backreaction}
\label{IR-UV-IR background solution}
We consider the IR-UV-IR warped geometry compactified on an interval,
$-L\leq y\leq L$, where a UV-brane with positive tension is located at $y=0$, and two negative tension IR-branes are located at $y=\pm L$. Note that the end points of the interval at $y=\pm L$ are not the fixed points of the $\mathbb{Z}_2$, the only fixed point
is at $y=0$, which is different from the $S_1/\mathbb{Z}_2$ orbifold (RS1 geometry) where $y=0$ and $y=L$ are both fixed points of the $\mathbb{Z}_2$.
The 5D gravity action for such a geometry, without taking into account any backreaction due to the presence of matter content, can be written as~\footnote{We use the metric signature $(-,+,+,+,+)$
and the unit system such that the 5D Planck mass $M_\ast=1$.},
\beq
S_G=\int d^5x\sqrt{-g}\left\{\frac R2-\Lambda_B-\lambda_{UV}\delta(y)-\lambda_{IR}\delta(y+L)-\lambda_{IR}\delta(y-L)\right\}+S_{GH},
\label{gravity_action}
\eeq
where $R$ is the Ricci scalar, $\Lambda_B$ is the bulk cosmological constant and $\lambda_{UV}(\lambda_{IR})$ are the brane
tensions at the UV(IR) fixed points. Above and henceforth the Dirac delta functions at $y=\pm L$ are defined in such a way that their integral is 1/2. Since our geometry is compact with boundaries, the action contains the Gibbons-Hawking boundary term, \footnote{The Gibbons--Hawking boundary term is needed in order to cancel variation of the Ricci scalar at the boundaries so that the RS metric \eqref{metric} is indeed a solution of the Einstein equations of motion.}
\beq
S_{GH}=-\int_{\partial{\cal M}}d^4x\sqrt{-\hat g} {\cal K},	\label{GH_term}
\eeq
where ${\cal K}$ is the intrinsic curvature of the surface of the boundary manifold $\partial{\cal M}$, given by
\beq
{\cal K}=-\hat g^{\mu\nu}\nabla_\mu n_\nu=\hat g^{\mu\nu}\Gamma^M_{\mu\nu}n_M,
\eeq
with $n_M$ being the unit normal vector to the surface of the boundary manifold $\partial{\cal M}$ and $\hat g_{\mu\nu}$ is the
induced boundary metric. For the 5D manifold with 4D Poincar\'e invariance ($n^5=1$ and $n^\mu=0$), the intrinsic curvature reduces to
\beq
{\cal K}=-\frac12\hat g^{\mu\nu}\partial_5\hat g_{\mu\nu}.
\eeq
The solution of the Einstein equations resulting from the above action is the RS metric \eqref{metric}, where the AdS curvature $k$ is related to $\Lambda_B$ by
\beq
\Lambda_B=-6k^2.    \label{lambda_k}
\eeq

Since the above setup is compactified on an interval $y\in [-L,L]$, rather than on a circle as in RS1, one needs
to be careful and show that the solution \eqref{metric} is compatible with the boundaries and that the effective 4D
cosmological constant is zero, see also \cite{Lalak:2001fd}. We will see below that we need a {\it fine tuning} between the 5D cosmological constant
$\Lambda_B$ and the brane tensions $\lambda_{UV,IR}$ in order to get zero 4D cosmological constant. One can calculate
the  effective 4D cosmological constant $\Lambda_{4}$ from the action \eqref{gravity_action} by integrating out
the extra-dimension,
\beq
\Lambda_4=-\int_{-L}^L dy\sqrt{-g}\left\{\frac R2-\Lambda_B-\lambda_{UV}\delta(y)-\lambda_{IR}\big[\delta(y+L)+\delta(y-L)\big]\right\}+\sqrt{-\hat g} {\cal K}\Big\vert_{-L}^{L}, \label{Lambda_4_a}
\eeq
where $R=-20 A^{\p2}-8A^{\p\p}$ and $\Lambda_B=-6A^{\p2}$ corresponding to the solution \eqref{metric}. Using $A(y)=-k|y|$ we find,
\beq
\Lambda_4=\left(\lambda_{UV}-6k\right)+\left(\lambda_{IR}+6k\right)e^{-4kL},          \label{Lambda_4_b}
\eeq
which can only be zero if
\beq
\lambda_{UV}=- \lambda_{IR}=6k.
\eeq
This result explicitly shows that one needs a positive tension brane at $y=0$ and two negative tension branes at $y=\pm L$ in order to obtain zero 4D cosmological constant. This is the usual fine tuning which appears in
brane world scenarios \cite{Randall:1999ee,Goldberger:1999uk,DeWolfe:1999cp}. Hence we have a 5D geometry with AdS
solution \eqref{metric} with negative bulk cosmological constant and a positive tension brane in the middle and two equal negative
tension branes at the end of the interval, see Fig. \ref{IR-UV-IR}.

We would like to mention here that we are considering a rigid IR-UV-IR geometry where the distance $L$ is tuned in order to solve the hierarchy problem. To stabilize the IR-UV-IR setup one may consider a mechanism like the one proposed by Goldberger and Wise (GW) \cite{Goldberger:1999uk,DeWolfe:1999cp} introducing a bulk scalar field with appropriate brane potentials such that the energy minimization would set the size of the 5D interval and yields a compactification scale that would solve the hierarchy problem. Since one of our aims is to analyze EWSB due to a 5D $SU(2)$ Higgs doublet in the IR-UV-IR model, therefore it is natural to consider the bulk $SU(2)$ Higgs doublet as the GW stabilizing field. We explore this option in the following subsection. A similar analysis for the case of RS1 has been considered in Ref.~\cite{Geller:2013cfa}. However, there the full scalar-gravity coupled equations are not solved,
the authors adopt a small backreaction anstaz. In this work, we are using so-called {\it superpotential method} to solve the full scalar-gravity coupled equations analytically.

\subsection{The IR-UV-IR model: with backreaction}
\label{The IR-UV-IR model: with backreaction}
In this subsection we employ an $SU(2)$ Higgs doublet in the bulk of IR-UV-IR model and obtain the background solutions for the 5D scalar-gravity coupled theory. Although our solution generating technique is adopted for a specific geometric configuration (IR-UV-IR model), this approach can be used to solve the Higgs-gravity backreaction in any warped extra-dimensional model (e.g. RS1) with a bulk Higgs. We use the following most general 4D Poincar\'e invariant metric ansatz:
\beq
ds^2=e^{2A(y)}\eta_{\mu\nu}dx^\mu dx^\nu+dy^2, \label{metric_sg}
\eeq
where $A(y)$ is a general $y$-dependent warp-function and $\eta_{\mu\nu}$ represents the 4D Minkowski metric. We consider the following scalar-gravity action for our model,
\begin{align}
S_{SG}=\int d^5x \sqrt{-g}&\Big\{\frac{R}{2}-\left\vert D_M H\right\vert^2 -V_{B}(H)	\notag\\
&-V_{UV}(H)\delta(y)-V_{IR}(H)\Big[\delta(y+L)+\delta(y-L)\Big]\Big\}+S_{GH},    \label{action_sg}
\end{align}
where $R$ is the 5D Ricci scalar and $H$ is the $SU(2)$ Higgs doublet. Whereas, $V_{B}(H)$ and $V_{UV(IR)}(H)$ are the bulk and UV(IR) brane potentials, respectively.
Above, the $S_{GH}$ is the Gibbons-Hawking boundary action defined in \eq\eqref{GH_term}.

We can write the $y$-dependent vacuum expectation value (vev) of the $SU(2)$ Higgs doublet as:
\beq
\langle H\rangle=\frac{1}{\sqrt2}\bpm 0\\ \phi(y) \epm.	\label{higgs_vev}
\eeq
In order to find the background solutions we need to solve the coupled scalar-gravity Einstein equations. The equations for the background fields $A(y)$ and $\phi(y)$ following from the above action \eqref{action_sg} and the metric ansatz \eqref{metric_sg} can be written as,
\begin{align}
6A^{\prime2}&=\frac{1}{2}\phi^{\prime2}-V_B(\phi),\label{eom01}\\
3A^{\prime\prime}+6A^{\prime2}&=-\frac{1}{2}\phi^{\prime2}-V_B(\phi)-V_{UV}(\phi)\delta(y)-V_{IR}(\phi)\Big[\delta(y+L)+\delta(y-L)\Big],\label{eom02}\\
\phi^{\prime\prime}+4A^{\prime}\phi^{\prime}&=\frac{\partial V_B(\phi)}{\partial \phi}+\frac{\partial V_{UV}(\phi)}{\partial \phi}\delta(y)+\frac{\partial V_{IR}(\phi)}{\partial \phi}\Big[\delta(y+L)+\delta(y-L)\Big]. \label{eom03}
\end{align}
\paragraph{Superpotential method:} In the following we will layout the so-called {\it superpotential method} for solving the above set of coupled scalar-gravity equations \cite{DeWolfe:1999cp}. Although the use of this method is motivated by supersymmetry, no supersymmetry is involved in
our setup. The method is elegant and
very efficient, in particular it applies to the system of second order differential equations
\eqref{eom01}-\eqref{eom03} and reduces them to a set of first order ordinary differential equations which are much easier to deal with. It is assumed that the scalar potential $V_B(\phi)$ could be expressed
in terms of the superpotential $W(\phi)$ as \cite{DeWolfe:1999cp,Csaki:2000zn},
\begin{equation}
V_B(\phi)=\frac{1}{8}\left( \frac{\partial W(\phi)}{\partial \phi}\right)^{2}-\frac{1}{6}W(\phi)^{2},
\label{potential}
\end{equation}
where the superpotential $W(\phi)$ satisfies,
\begin{align}
 \phi^{\prime}&=\frac12\frac{\partial W(\phi)}{\partial \phi},  &A^{\prime}&=-\frac{1}{6} W(\phi),
 \label{super_potential_eqs}
 \end{align}
 along with the following jump at $y=0$ and boundary conditions $y=\pm L$:
 \begin{align}
\frac12\Big[W(\phi)\Big]_0&=V_{UV}(\phi)\Big\vert_{\phi=\phi(0)}, &\frac12\Big[\frac{\partial W(\phi)}{\partial \phi}\Big]_0&=\frac{\partial V_{UV}(\phi)}{\partial \phi}\Big\vert_{\phi=\phi(0)}.	\label{jump_conditions_W}\\
W(\phi)\Big\vert_{\pm L}&=-V_{IR}(\phi)\Big\vert_{\phi=\phi(\pm L)}, &\frac{\partial W(\phi)}{\partial \phi}\Big\vert_{\pm L}&=-\frac{\partial V_{IR}(\phi)}{\partial \phi}\Big\vert_{\phi=\phi(\pm L)}.	\label{boundary_conditions_W}
\end{align}
Above, the jump across the UV-brane of $j(y)=W(\phi) \text{ and }\partial W(\phi)/\partial \phi$, is defined as follows:
\beq
\big[j(y)\big]_{0}\equiv\lim_{\epsilon\to0}\big\{j(0+\epsilon)-j(0-\epsilon)\big\}. \label{jump}
\eeq

\begin{figure}
\begin{center}
\includegraphics[width=0.45\textwidth]{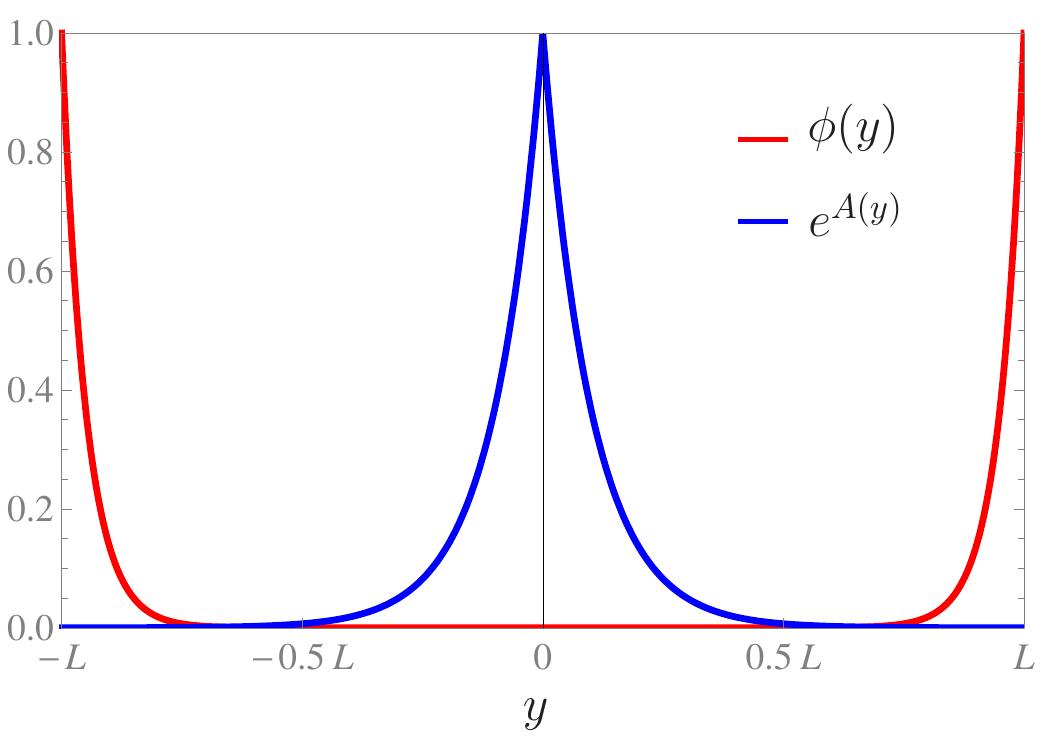}\hsp
\includegraphics[width=0.45\textwidth]{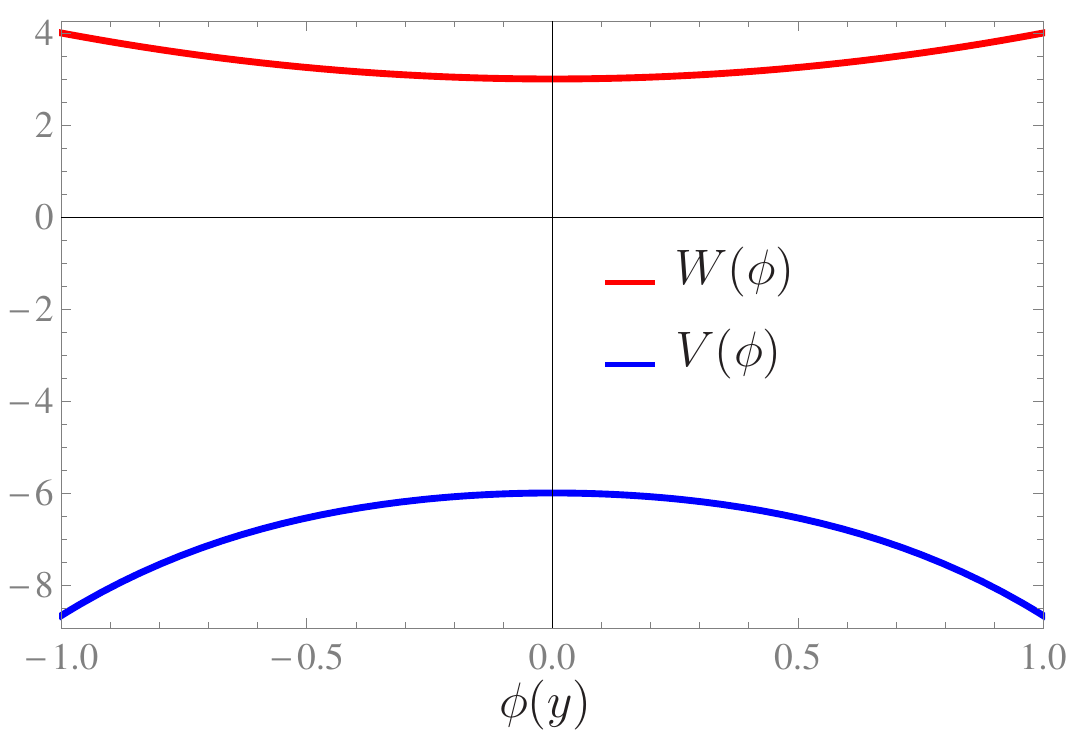}
\end{center}
\caption{The left graph shows the $y$-dependent background vev $\phi(y)$ and the warp factor $e^{A(y)}$ as function of $y$, while the right graph illustrates the shapes of superpotential $W(\phi)$ and the bulk scalar potential $V_B(\phi)$ as a function of background vev $\phi$. The parameter choice adopted for the graphs is: $\beta=0$, $k=1$ and $\phi_{IR}=1$.}
\label{phiAWV}
\end{figure}

Let us consider the following form of the superpotential $W(\phi)$
\beq
W(\phi)=\left\{\begin{array}{ll} \enspace\;6k+(2+\beta)k\phi^2 &\lsp\text{for}\lsp  0<y<L\\
-6k-(2+\beta)k\phi^2 &\lsp\text{for}\hsp -L<y<0\end{array}\right., \label{superpotential}
\eeq
where $\beta\equiv\sqrt{4+\mu_B^2/k^2}$ parameterises the bulk mass $\mu_B$ of the Higgs field and $k$ is a constant of order $M_\ast$. The above form  of $W(\phi)$ is $\mathbb{Z}_2$ odd under $y\to-y$, which ensures, by the virtue of \eq\eqref{super_potential_eqs}, that $A(y)$ and $\phi(y)$ are $\mathbb{Z}_2$ even. We get the scalar potential $V_B(\phi)$ from \eq\eqref{potential} as
\beq
V_B(\phi)=-6k^2+\frac12\mu_B^2\phi^2-\frac{k^2}2(2+\beta)^2\phi^4. \label{bulk_potenial}
\eeq
We employ the following forms of the brane-localized potentials,
\begin{align}
V_{UV}(\phi)&=W(\phi)+\frac{\lambda_{UV}}{4k^2}\big(\phi^2-\phi_{UV}^2\big)^2,	\label{Vuv}\\
V_{IR}(\phi)&=-W(\phi)+\frac{\lambda_{IR}}{4k^2}\big(\phi^2-\phi_{IR}^2\big)^2,	\label{Vir}
\end{align}
where $\phi_{UV(IR)}$ is a constant value of the background vev at $y=0 (\pm L)$ and $\lambda_{UV(IR)}$ is the quartic coupling at the UV(IR) brane.
The background vev $\phi(y)$ and the warp-function $A(y)$ can be obtained by integrating the first order equations \eqref{super_potential_eqs} as
\begin{align}
\phi(y)&=\phi_{IR}e^{(2+\beta)k(|y|-L)},      \label{scalar_phi}\\
A(y)&=-k|y| -\frac{1}{12}\phi_{IR}^2e^{-2(2+\beta)kL}\Big[e^{2(2+\beta)k|y|}-1\Big].  \label{warp_function_A}
\end{align}
Moreover, we impose the following normalization condition for the background vev:
\beq
\int_{-L}^{L}dy e^{2A(y)}\phi^2(y)=v_{SM}^2,       \label{phi_vev_norm}
\eeq
where $v_{SM}$ is the SM vev.
The $\phi_{IR}$ resulting from the above normalization reads (see also \eq\eqref{vp_sol_toy}):
\beq
\phi_{IR}=v_{SM}\sqrt{k(1+\beta)}e^{kL},	\label{phi_ir}
\eeq
As we will see below in order to solve the gauge hierarchy problem, one needs $kL\simeq 37$ and for $\phi_{IR}\sim \co(\mpl^{3/2})$, $k\approx \co(\mpl)$ and $\beta\approx\co(1)$, the above expression implies that then $v_{SM}\sim \co(\text{TeV})$.
In Figure \ref{phiAWV}, we have plotted the $y$-dependent background vev $\phi(y)$ and the warp factor $e^{A(y)}$ as function of $y$ in the left panel, while the right panel shows the superpotential $W(\phi)$ and the bulk scalar potential $V_B(\phi)$ as a function of $\phi$.

In order to gain some intuition, it is instructive to count number of parameters and constraints to see how many fine-tunings are necessary to obtain solutions with Minkowski D3-branes. There are three integration constants coming from the first order Eqs.~\eqref{potential}
and \eqref{super_potential_eqs}. One of these three, $A(0)$, is trivial additive constant hence we are left with two parameters to be fixed. Let us count the number of constraints: there are two constraints from the jump at $y=0$ from \eq\eqref{jump_conditions_W} and two constraints from the boundaries at $\pm L$ Eq.~\eqref{boundary_conditions_W} (note that the boundary conditions are same at $\pm L$). We can integrate \eq\eqref{potential} to fix the integration constant by requiring $W(\phi_{UV})=V_{UV}(\phi_{UV})$. This requirement together with \eq\eqref{potential} will trivially satisfy the second jump condition of \eq\eqref{jump_conditions_W}. We can then integrate the first equation of \eq\eqref{super_potential_eqs} to get $\phi(y)$ and the position of the IR-brane is fixed w.r.t. the UV-brane by requiring $\phi(L)=\phi_{IR}$. This fixes all the parameters of the theory but we are still left with two constraints \eq\eqref{boundary_conditions_W} which are not independent of each other as fixing $W(\phi_{IR})=-V_{IR}(\phi_{IR})$ will trivially satisfy second equation of \eq\eqref{boundary_conditions_W}. Hence we are left with precisely one fine-tuning which is required to ensure flat-brane solutions and is a common pathology of RS-like models, see also Ref.~\cite{DeWolfe:1999cp}.

Let us check that the 4D cosmological constant $\Lambda_4$ is indeed zero after having this fine-tuning is achieved.
The 4D cosmological constant for this scalar-gravity theory \eqref{action_sg} is given by the following integral calculated for the background solutions
\begin{align}
\Lambda_4&=-\int_{-L}^L dy\sqrt{-g}\left\{\frac R2-\frac12 \phi^{\p2}-V_B(\phi)-V_{UV}(\phi)\delta(y)-V_{IR}(\phi)\big[\delta(y+L)+\delta(y-L)\big]\right\}+\sqrt{- \hat g} {\cal K}\Big\vert_{-L}^{L} \notag\\
&=-\int_{-L}^L dye^{4A}\bigg\{6\left(A^\p+\frac16W\right)^2-\frac12\left(\phi^\p-\frac12\frac{\partial W}{\partial\phi}\right)^2-e^{-4A}\frac{d}{dy}\left[e^{4A}\left(4A^\p+\frac12W\right)\right]	\notag\\
&\Lsp\hsp+\frac12\frac{\partial W}{\partial y}-V_{UV}(\phi)\delta(y)-V_{IR}(\phi)\big[\delta(y+L)+\delta(y-L)\big]\bigg\}-4e^{4A}A^\p\Big\vert_{-L}^{L} \notag\\
&=-\int_{-L}^L dye^{4A}\bigg\{6\left(A^\p+\frac16W\right)^2-\frac12\left(\phi^\p-\frac12\frac{\partial W}{\partial\phi}\right)^2\bigg\}	\notag\\
&\hsp-\frac12\big[ W\big]_0+V_{UV}[\phi(0)]+\frac{e^{4A}}2W\Big\vert_{-L}^L+\frac{e^{4A(L)}}2V_{IR}[\phi(L)]+\frac{e^{4A(L)}}2V_{IR}[\phi(-L)].	\label{4d_cc}
\end{align}
Above, in the second line we have only used \eq\eqref{potential} and the value of the intrinsic curvature of the boundary manifolds ${\cal K}$. In the last line we have adopted the fact that the superpotential $W(\phi)$ \eqref{superpotential} is discontinuous at $y=0$, hence the jump at $y=0$ follows from the total derivative term. Now it is straightforward to see in the above equation that the full squares vanish due to \eq\eqref{super_potential_eqs}, whereas the last line vanishes due to the jump and boundary conditions \eqref{jump_conditions_W}-\eqref{boundary_conditions_W}. Hence we conclude that the 4D cosmological constant on the branes is indeed zero, as anticipated from our metric ansatz \eqref{metric_sg}.

The brane separation $L$ can be determined by \eq\eqref{scalar_phi} as:
\beq
kL=\frac{1}{2+\beta}\ln\left(\frac{\phi_{IR}}{\phi_{UV}}\right).
\eeq
In order to solve the gauge hierarchy problem, the number of $e$-foldings is required to be as follows
\beq
A(0)-A(L)\simeq \ln\left(\frac{\mpl}{m_{EW}}\right)\approx 37.  \label{e_folding}
\eeq
By using the \eq\eqref{warp_function_A}, we get:
\begin{align}
A(0)-A(L)
&= \frac{1}{2+\beta}\ln\left(\frac{\phi_{IR}}{\phi_{UV}}\right)+\frac{1}{12}\big(\phi_{IR}^2-\phi_{UV}^2\big).  \label{e_folding}
\end{align}
Note that for $\beta\geq0$,  both terms in the above relation can contribute to give the desired number of $e$-foldings.
When the first term contributes mostly ({\it weak backreaction}), so $\phi_{IR}\gg\phi_{UV}$, then  $kL\simeq37$.
Whereas, when the major contribution to the number of $e$-foldings comes from the second term ({\it strong backreaction}), i.e. $\phi_{UV}\sim\co(1)$ and $\phi_{IR}\sim\co(20)$
then $kL\sim \co(1)$.
In this paper we consider the weak backreaction scenario, i.e. $kL\simeq37$, which implies that the second term in the warp-function \eqref{warp_function_A} is negligible and hence the form of warp-function is $A(y)\simeq -k|y|$ which is the same as in the previous subsection with no backreaction. Note that the weak backreaction scenario is also required in order to insure the SM vev $v_{SM}\simeq246\gev$ below the KK-scale, see e.g. Ref.~\cite{Geller:2013cfa}. The case of strong backreaction is also interesting as the fine-tuning required is much less than that of the weak backreaction scenario but we will not consider it here.


\subsection{Warped KK-Parity}
\label{KK-Parity}

\noindent In this section we employ the background solution for the $\mathbb{Z}_2$ symmetric background (IR-UV-IR) geometry considered in Sec.~\ref{IR-UV-IR background solution} and show how KK-parity is manifested within this geometric setup. The IR-UV-IR geometry of Sec.~\ref{IR-UV-IR background solution} is $\mathbb{Z}_2$-symmetric  and we will consider this symmetry to be exact for our 5D theory. If the 5D theory has this $\mathbb{Z}_2$-parity (symmetry) then the Schr\"odinger-like potential for all the fields is symmetric, resulting in even (symmetric) and odd (antisymmetric) eigenmodes.  Thus, a general field $\Phi(x,y)$ can be KK decomposed
as follows
\beq
\Phi(x,y)=\sum_{n}\phi_n(x)f_n(y),
\eeq
where, due to the $\mathbb{Z}_2$ geometry, the wave functions $f_n(y)$ are either even or odd, so that
\beq
\Phi(x,y)\equiv\Phi^{(\pm)}(x,y),
\eeq
with
\begin{align}
\Phi^{(+)}(x,y)&=\sum_n\phi^{(+)}_n(x)f^{(+)}_n(y)\xrightarrow{y\to-y}+\Phi^{(+)}(x,y),   \\
\Phi^{(-)}(x,y)&=\sum_n\phi^{(-)}_n(x)f^{(-)}_n(y)\xrightarrow{y\to-y}-\Phi^{(-)}(x,y).
\end{align}
Due to the geometric $\mathbb{Z}_2$ symmetry, a single odd KK-mode cannot couple to two even KK-modes in the 4D effective theory,
therefore the lowest odd KK-mode will be stable and may serve as a dark matter candidate.

Furthermore, as the geometry is $\mathbb{Z}_2$ symmetric in $y\in[-L,L]$, the continuity conditions for odd and even modes at $y=0$ strongly impact the physics scenario.  Our choice will be that the odd (even) modes
satisfy Dirichlet (Neumann or mixed)  boundary (jump) conditions (b.c.) at $y=0$, respectively. As for the odd modes, continuity implies that they must be zero at $y=0$, but we could also have demanded the Neumann conditions that their $y$ derivative be zero at $y=0$.  We choose not to impose this additional b.c. in this work.  As regards the even modes, one cannot choose Dirichlet b.c. at $y=0$ because of the presence of the UV-brane and associated ``jump" conditions following from the equations of motion.

\section{SM EWSB by bulk Higgs doublet -- the truncated-inert-doublet model}
\label{SM EWSB due to bulk Higgs doublet}
\noindent In this section we consider all the SM fields in the bulk and study phenomenological implications of our
symmetric geometry. Hereafter, we consider only the weak backreaction scenario discussed in Sec.~\ref{The IR-UV-IR model: with backreaction} and we employ the metric \eqref{metric} as a solution of the IR-UV-IR geometric background. Note also that vev of effective 4D Higgs field is of the electroweak scale which is much smaller than the gravity (Planck mass) scale (see \eq\eqref{phi_ir} and discussion below), therefore their back-reaction on the background geometry would be negligible, see sub-Sec.~\ref{The IR-UV-IR model: with backreaction} and Refs.~\cite{Geller:2013cfa,Cox:2013rva}. Moreover, we follow closely the Abelian case~\ref{EWSB by vacuum expectation value of KK modes} for EWSB of the 5D SM gauge group and as shown in Appendix~\ref{SSB in the IR-UV-IR model: the Abelian Higgs mechanism} this approach is equivalent, at the zero-mode level, to the canonical approach (Appendix~\ref{EWSB by vacuum expectation value of 5D Higgs field}), i.e. when the bulk Higgs field is expanded around the background vev.

The 5D action for the electroweak sector of the SM can be written as~\footnote{Note that the Higgs part of the Lagrangian is not exactly same as considered in Sec.~\ref{The IR-UV-IR model: with backreaction} for the analysis of the backreaction, here we neglect the quartic terms in the bulk and UV-brane potentials solely for phenomenological convenience since these terms are suppressed, see the main text below. Moreover, we introduce some convenient and familiar notations for brane potential parameters.}
\begin{align}
S=-\int d^5x \sqrt{-g}\bigg\{&\frac14 F^{a}_{MN}F^{aMN}+\frac14 B_{MN}B^{MN}+\left\vert D_M H\right\vert^2 +\mu_B^2|H|^2   \notag\\
&+V_{IR}(H)\delta(y+L)+V_{UV}(H)\delta(y)+V_{IR}(H)\delta(y-L)\bigg\},    \label{action_SM}
\end{align}
where $F^a_{MN}$ and $B_{MN}$ are the 5D field strength tensors for $SU(2)$ and $U(1)_Y$, respectively with $a$
being the number generators of $SU(2)$. Above, $H$ is the $SU(2)$ doublet and its brane potentials are
\beq
V_{UV}(H)=\frac{m^2_{UV}}{k}|H|^2,   \Lsp V_{IR}(H)=-\frac{m^2_{IR}}{k}|H|^2+\frac{\lambda_{IR}}{k^2}|H|^4.
\label{boudary_potentials}
\eeq
In our approach, we do not put the Higgs quartic terms in the bulk nor on the UV-brane since we want EWSB to take place near the IR-brane.  The covariant derivative $D_M$ is defined as follows:
\beq
D_M =\partial_M-i\frac{g_5}{2} \tau^aA^a_M -i\frac{g_5^\p}{2} B_M,      \label{co_dir_M_SM}
\eeq
where $\tau^a$ are Pauli matrices and $g_5(g^\p_5)$ is the coupling constant for the $A_M^a(B_M)$ fields.

There is an important comment in order here concerning the above particular forms of bulk and brane-localized potentials. We have dropped quartic terms in the UV and the bulk potentials, even though there is no symmetry that would protect this choice. However there are phenomenological arguments that support such an option. As one can notice the UV Higgs quartic operator, i.e. $V_{UV}(H)\supset\lambda_{UV}/k^2 |H|^4$ is highly suppressed as $\lambda_{UV}/k^2\sim {\cal O}(M_{\text{Pl}}^{-2})$. Whereas for the IR Higgs quartic operator, suppression of  $\lambda_{IR}/k^2$ is reduced to $\sim{\cal O}(m_{KK}^{-2})$ due to the non-trivial warp factor at the IR-brane, see also Refs.~\cite{Archer:2012qa,Archer:2014jca}. Similarly, the bulk quartic term would also be suppressed by some intermediate scale. For simplicity we ignore those terms.
We shell emphasize that the reasoning behind ignoring the bulk and the UV-brane localized quartic terms is purely phenomenological in its nature.
It is also worth recalling that in this work we are going to investigate the possibility of providing a dark matter candidate as a lowest mass odd scalar KK mode.
It is fair to expect that those results will not receive any substantial corrections in the presence of quartic UV and brane terms.
Besides that, this kind of phenomenological approximation of ignoring the bulk and UV-brane quartic terms is widely used in the literature on bulk Higgs scenarios in RS1-like models, see for example \cite{Cacciapaglia:2006mz,Falkowski:2008fz,Azatov:2009na,Cabrer:2010si,Cabrer:2011fb,Cabrer:2011vu,Geller:2013cfa,Archer:2012qa,Frank:2013un,Malm:2013jia,Cox:2013rva,Archer:2014jca,Agashe:2014jca}.

Note also that due to the $\mathbb{Z}_2$ geometric symmetry, physics of the full IR-UV-IR setup can be described completely by a single copy of RS1, i.e. UV-IR setup, however in that scenario each bulk field would be a subject of Neumann (or mixed) and Dirichlet boundary conditions at $y=0$ for even and odd fields, respectively.

It is instructive to make the usual redefinition of the gauge fields,
\begin{align}
W^{\pm}_M(x,y)&\equiv\frac{1}{\sqrt2}\Big(A^1_M\mp iA^2_M\Big),    \\
Z_M(x,y)&\equiv\frac{1}{\sqrt{g^2_5+g^{\p2}_5}}\Big(g_5A^3_M-g^\p_5B_M\Big),\notag\\
A_M(x,y)&\equiv\frac{1}{\sqrt{g^2_5+g^{\p2}_5}}\Big(g^\p_5A^3_M+g_5B_M\Big).
\end{align}
Analogous to the 4D procedure, we define the 5D Weinberg angle $\theta$
as follows:
\beq
\cos\theta\equiv \frac{g_5}{\sqrt{g^2_5+g^{\p2}_5}}, \lsp
\sin\theta\equiv \frac{g^\p_5}{\sqrt{g^2_5+g^{\p2}_5}}. \label{sin_cos}
\eeq
The 5D gauge fields corresponding to the gauge group $SU(2)\times U(1)_Y$ are then
\beq
\mathbb{A}_M(x,y)\equiv \bpm \sin\theta A_M+\frac{\cos^2\theta-\sin^2\theta}{2\cos\theta}Z_M & \frac{1}{\sqrt2}W^+_M\\
\frac{1}{\sqrt2}W^-_M& -\frac{1}{2\cos\theta}Z_M \epm.      \label{A_matrix}
\eeq
The gauge transformations for the Higgs doublet $H(x,y)$ and gauge matrix $\mathbb{A}_M$ under the gauge group
$SU(2)\times U(1)_Y$ can be written as
\begin{align}
H(x,y)\to H^\p(x,y)&=U(x,y)H(x,y),      \label{H_gauge_trans}\\
\mathbb{A}_M(x,y)\to \mathbb{A}_M^\p(x,y)&=U(x,y)\mathbb{A}_M(x,y)U^{-1}(x,y)-\frac{i}{g_5}(\partial_M U(x,y)) U^{-1}(x,y),
\label{A_gauge_trans}
\end{align}
where $U(x,y)$ is the unitary matrix corresponding to the fundamental representation of $SU(2)\times U(1)_Y$
gauge transformations.

We will choose the 5D axial gauge analogous to the Abelian case~\ref{EWSB by vacuum expectation value of KK modes} by taking $\mathbb{A}_5(x,y)=0$. Note that we can always find $U(x,y)$
such that the axial gauge is manifest, i.e. $\mathbb{A}_5(x,y)=0$. We employ an axial gauge choice for  the
non-Abelian case of the form
\begin{align}
U(x,y)=\widehat U(x){\cal P}e^{-ig_5\int_0^ydy^\p \mathbb{A}_5(x,y^\p)},
\label{axial_gauge_cond}
\end{align}
where $\widehat U(x)$ is the residual 4D gauge transformation and ${\cal P}$ denotes path-ordering
of the exponential. Another key point for the later discussion is that this 4D residual gauge
transformation $\widehat U(x)$ is independent of $y$ and thus even under the geometric parity.

As we have demonstrated in Appendix~\ref{SSB in the IR-UV-IR model: the Abelian Higgs mechanism}, due to the
symmetric geometry the background fields in the IR-UV-IR setup separate into even and odd bulk
wave functions. Hence, it is straightforward to
generalize the results obtained in Appendix~\ref{EWSB by vacuum expectation value of KK modes} for the
Abelian model to the electroweak sector of the SM. Let us start by decomposing the Higgs doublet and gauge
fields into components of definite parity as follows:
\beq
H(x,y)=H^{(+)}(x,y)+H^{(-)}(x,y),      \lsp V_M(x,y)=V^{(+)}_{M}(x,y)+V^{(-)}_{M}(x,y),     \label{even_odd_H_A_SM}
\eeq
where $V_M\equiv (A_M, W^{\pm}_M, Z_M)$.
We can write the action \eqref{action_SM} up to quadratic level in the $\mathbb{A}_5(x,y)=0$ gauge as
\begin{align}
S^{(2)}=&-\int d^5x \sqrt{-g}\bigg\{\frac12 {\cal W}^{+}_{(+)\mu\nu}{\cal W}_{(+)}^{-\mu\nu}+ \partial_5W^{+}_{(+)\mu}\partial^5W_{(+)}^{-\mu}+\frac14 {\cal Z}^{(+)}_{\mu\nu}{\cal Z}_{(+)}^{\mu\nu}+\frac12 \partial_5Z^{(+)}_{\mu}\partial^5Z_{(+)}^{\mu}\notag\\
&+\frac12 {\cal W}^{+}_{(-)\mu\nu}{\cal W}_{(-)}^{-\mu\nu}+\partial_5W^{+}_{(-)\mu}\partial^5W_{(-)}^{-\mu}+\frac14 {\cal Z}^{(-)}_{\mu\nu}{\cal Z}_{(-)}^{\mu\nu}+\frac12 \partial_5Z^{(-)}_{\mu}\partial^5Z_{(-)}^{\mu}\notag\\
&+\frac14 {\cal F}^{(+)}_{\mu\nu}{\cal F}_{(+)}^{\mu\nu}+\frac12 \partial_5A^{(+)}_{\mu}\partial_5A_{(+)}^{\mu}  +\frac14 {\cal F}^{(-)}_{\mu\nu}{\cal F}_{(-)}^{\mu\nu}+\frac12 \partial_5A^{(-)}_{\mu}\partial^5A_{(-)}^{\mu}   \notag\\
&+ \mathbb{D}_M H^{(+)\dag}\mathbb{D}^MH^{(+)}  +\mu_B^2|H^{(+)}|^2 + \mathbb{D}_M H^{(-)\dag}\mathbb{D}^MH^{(-)} +\mu_B^2|H^{(-)}|^2 \notag\\
&+\frac{m^2_{UV}}{k}|H^{(+)}|^2\delta(y)-\frac{m^2_{IR}}{k}\big(|H^{(+)}|^2+|H^{(-)}|^2\big)\big[\delta(y+L)+\delta(y-L)\big]\bigg\},    \label{action_5d_sm}
\end{align}
where we have adopted the following definitions:
\begin{align}
{\cal \tilde V}^{(\pm)}_{\mu\nu}\equiv \partial_\mu \tilde V^{(\pm)}_{\nu}&-\partial_\nu \tilde V^{(\pm)}_{\mu}, \lsp {\cal F}^{(\pm)}_{\mu\nu}\equiv \partial_\mu A^{(\pm)}_{\nu}-\partial_\nu A^{(\pm)}_{\mu},    \label{photon_emt}\\
\mathbb{D}_\mu \bpm H^{(+)} \\ H^{(-)}\epm&\equiv\left[\partial_\mu  -ig_5\bpm \mathbb{A}^{(+)}_\mu & \mathbb{A}^{(-)}_\mu \\
\mathbb{A}^{(-)}_\mu &\mathbb{A}^{(+)}_\mu \epm \right] \bpm H^{(+)} \\ H^{(-)}\epm,   \label{c_dir_mu_pm}\\
\mathbb{D}_5 \bpm H^{(+)} \\ H^{(-)}\epm&\equiv \left[ \partial_5-ig_5\bpm \mathbb{A}^{(-)}_5 & \mathbb{A}^{(+)}_5 \\
\mathbb{A}^{(+)}_5 &\mathbb{A}^{(-)}_5 \epm\right] \bpm H^{(+)} \\ H^{(-)}\epm,   \label{c_dir_5_pm}
\end{align}
where $\tilde V_\mu\equiv(W^\pm_\mu, Z_\mu)$ and $\mathbb{A}_M^{(\pm)}$ was defined in \eqref{A_matrix}.

It is convenient to write the Higgs doublets in the following form:
\beq
\bpm H^{(+)} \\ H^{(-)}\epm=e^{ig_5 (\Pi^{(+)}\mathds{1}+\Pi^{(-)}\tau_1)} \bpm {\cal H}^{(+)} \\ {\cal H}^{(-)}\epm,   \label{higgs_redef}
\eeq
where ${\cal H}$ and $\Pi$ are defined as (the parity indices are suppressed)
\begin{align}
{\cal H}(x,y)&\equiv \frac{1}{\sqrt2}\bpm 0\\ h(x,y)\epm,         \label{hat_H}\\
\Pi(x,y)&\equiv \bpm \frac{\cos^2\theta-\sin^2\theta}{2\cos\theta}\pi_Z & \frac{1}{\sqrt2}\pi^+_W\\
\frac{1}{\sqrt2}\pi^-_W& -\frac{1}{2\cos\theta}\pi_Z \epm.      \label{pi_matrix}
\end{align}

We KK-decompose the Higgs doublets $H^{(\pm)}(x,y)$ and the gauge fields $V^{(\pm)}_\mu(x,y)$ as
\begin{align}
{\cal H}^{(\pm)}(x,y)&=\sum_n {\cal H}^{(\pm)}_n(x)f^{(\pm)}_n(y),   \label{KK_H_sm}\\
\pi^{(\pm)}_{\tilde V}(x,y)&=\sum_n \pi_{{\tilde V}n}^{(\pm)}(x)a^{(\pm)}_{\tilde Vn}(y),   \label{KK_pi_sm}\\
V^{(\pm)}_\mu(x,y)&=\sum_n V^{(\pm)}_{\mu n}(x)a^{(\pm)}_{V_n}(y),   \label{KK_A_sm}
\end{align}
where the wave-functions $f^{(\pm)}_n(y)$ and $a^{(\pm)}_{V_n}(y)$ satisfy
\begin{align}
-\partial_5 (e^{4A(y)}\partial_5 f^{(\pm)}_n(y))+\mu_B^2e^{4A(y)}f^{(\pm)}_n(y)&=m_n^{2(\pm)}e^{2A(y)}f^{(\pm)}_n(y),  \label{eom_fn_H}\\
-\partial_5 (e^{2A(y)}\partial_5 a^{(\pm)}_{V_n}(y))&=m_{V^{(\pm)}_n}^2a^{(\pm)}_{V_n}(y),  \label{eom_fn_A}
\end{align}
and, with our background geometry $A(y)=-k|y|$.
The $y$-dependent wave functions $f^{(\pm)}_n(y)$ and $a^{(\pm)}_{V_n}(y)$ satisfy the following
orthonormality conditions:
\begin{align}
\int_{-L}^{+L}dy e^{2A(y)}f^{(\pm)}_m(y)f^{(\pm)}_n(y)=\delta_{mn}, \lsp \int_{-L}^{+L}dy a^{(\pm)}_{V_m}(y)a^{(\pm)}_{V_n}(y)=\delta_{mn}. \label{norm_condition_HA_sm}
\end{align}
The even modes are subject to  jump conditions at $y=0$ while the odd modes are constrained by continuity at $y=0$, resulting in the following boundary conditions:
\begin{align}
\left(\partial_5 -\frac{m^2_{UV}}{k}\right)f^{(+)}_n(y)\Big\vert_{0}&=0,    \Lsp f^{(-)}_n(y)\Big\vert_{0}=0,  \label{H_bc0}\\
\partial_5a^{(+)}_{V_n}(y)\Big\vert_{0^+}&=0,    \Lsp a^{(-)}_{V_n}(y)\Big\vert_{0}=0.  \label{A_bc0}
\end{align}
The b.c. at $y=\pm L$ are:
\beq
\left(\pm\partial_5-\frac{m^2_{IR}}{k}\right)f^{(\pm)}_{n}(y)\Big\vert_{\pm L}=0,  \lsp \partial_5a^{(\pm)}_{V_n}(y)\Big\vert_{\pm L}=0.   \label{HA_bcL}
\eeq
As pointed out in the Abelian case~\ref{EWSB by vacuum expectation value of KK modes}, the choices of b.c. for $a^{(+)}_n(y)$ at $y=0,\pm L$ are motivated by the requirement that the even zero-mode profiles for gauge bosons be non-zero.

It is worth mentioning here that the choice of writing the Higgs doublets $H^{(\pm)}$ in the form of Eq. \eqref{higgs_redef} and using the KK decomposition
for the pseudoscalars $\pi^{(\pm)}_{\tilde V}$ as given in Eq.~\eqref{KK_pi_sm} are both motivated by model-building considerations discussed below.
The other possibility is to choose different KK bases and b.c. for the pseudoscalars $\pi^{(\pm)}_{\tilde V}$ such that after SSB these
pseudoscalars become Nambu-Goldstone bosons (NGB). The even zero-mode gauge bosons would then acquire masses by eating up the even-parity NGB,
whereas the odd-parity NGB would remain in the spectrum (the odd zero-mode gauge boson fields being zero, see below). An effective potential for the odd-parity NGB
would be generated through their interactions with gauge bosons, hence making them pseudo-NGB.
We don't follow this approach here but it is an interesting possibility in which the neutral odd pseudo-NGB would be a composite dark Higgs in the dual
CFT description.\footnote{At the final stages of the present
work, Ref.~\cite{Carmona:2015haa} appeared where the authors considered composite dark sectors. A similar construction can be naturally realized as a CFT dual to the model considered here.}

We assume that the KK-scale is high enough,    i.e.  $m_{KK}\sim\co(\text{few})\tev$,  that we can consider
an effective theory where only the lowest modes (zero-modes with masses much below $m_{KK}$) are kept.
It is important to note that the odd zero-mode wave functions obey  $a^{(-)}_{V_0}(y)=0$, as
can be easily seen from Eq. \eqref{eom_fn_A} along with the b.c. \eqref{A_bc0} and \eqref{HA_bcL}.
As a consequence of $a^{(-)}_{V_0}(y)=0$,   the odd zero-mode gauge fields $V^{(-)}_{0\mu}(x)$
and the odd Goldstone modes $\pi^{(-)}_{\tilde V_0}(x)$ will not be present in the effective 4D theory. Moreover, the even zero-mode gauge
profile is constant, i.e.  $a^{(+)}_{V_0}(y)=1/\sqrt{2L}$. Using the results from Appendix~\ref{SSB in the IR-UV-IR model: the Abelian Higgs mechanism}, we can determine  the values of the couplings and mass parameters in the effective 4D
theory in terms of the parameters of the fundamental 5D theory. The result is that we can write down the
effective 4D action for the zero-modes as
\begin{align}
S^{(2)}_{eff}=&-\int d^4x \bigg\{\frac14 {\cal F}^{0(+)}_{\mu\nu}{\cal F}_{0(+)}^{\mu\nu}+\frac14 {\cal Z}^{0(+)}_{\mu\nu}{\cal Z}_{0(+)}^{\mu\nu}+\frac12 {\cal W}^{+0(+)}_{\mu\nu}{\cal W}_{0(+)}^{-\mu\nu}+ \partial_\mu {\cal H}^{(+)\dag}_0\partial^\mu {\cal H}^{(+)}_0  \notag\\
&+ \partial_\mu {\cal H}^{(-)\dag}_0\partial^\mu {\cal H}^{(-)}_0 +m_0^{2(+)}|{\cal H}^{(+)}_0|^2 +m_0^{2(-)}|{\cal H}^{(-)}_0|^2 -ig_{4}\partial_\mu{\cal H}^{(+)\dag}_0\mathbb{M}^\mu{\cal H}^{(+)}_0 \notag\\
&+ig_4{\cal H}^{(+)\dag}_0\mathbb{M}_\mu^\dag\partial^\mu{\cal H}^{(+)}_0+g^{2}_{4}{\cal H}^{(+)\dag}_0\mathbb{M}_\mu^\dag\mathbb{M}^\mu{\cal H}^{(+)}_0   +g^{2}_{4}{\cal H}^{(-)\dag}_0\mathbb{M}_\mu^\dag\mathbb{M}^\mu{\cal H}^{(-)}_0\bigg\},    \label{eff_action_zero_0}
\end{align}
where $\mathbb{M}_\mu$ is defined as
\beq
\mathbb{M}_\mu\equiv\mathbb{U}^\dag\mathbb{\hat A}^{(+)}_{0\mu}\mathbb{U}+\frac{i}{g_4}\mathbb{U}^\dag\partial_\mu\mathbb{U}, \label{M_def}
\eeq
with $\mathbb{U}\equiv e^{ig_4 \widehat \Pi_0^{(+)}}$ and $g_4\equiv g_5/\sqrt{2L}$. In the above action ${\cal H}^{(\pm)}_0$ are real doublets defined in Eq. \eqref{hat_H}, implying that  ${\cal H}^{(\pm)\dag}_0={\cal H}^{(\pm)\intercal}_0$, whereas $\mathbb{\hat A}^{(+)}_{0\mu}$ and $\widehat \Pi_0^{(+)}$ are defined as (below we suppress the parity indices and zero-mode index):
\begin{align}
\mathbb{\hat A}_{\mu}(x)&\equiv \bpm \sin\theta A_{\mu}+\frac{\cos^2\theta-\sin^2\theta}{2\cos\theta}Z_{\mu} & \frac{1}{\sqrt2}W^+_{\mu}\\
\frac{1}{\sqrt2}W^-_{\mu}& -\frac{1}{2\cos\theta}Z_{\mu} \epm,      \label{A_matrix_0}\\
\widehat \Pi(x)&\equiv \bpm \frac{\cos^2\theta-\sin^2\theta}{2\cos\theta}\pi_{Z} & \frac{1}{\sqrt2}\pi^+_{W}\\
\frac{1}{\sqrt2}\pi^-_{W}& -\frac{1}{2\cos\theta}\pi_{Z} \epm.      \label{pi_matrix_0}
\end{align}
It is important to comment here that the above action is manifestly gauge invariant under the following
$SU(2)\times U(1)_Y$ gauge transformation,
\beq
\mathbb{\hat A}^{(+)}_{\mu}\to \widehat U\mathbb{\hat A}^{(+)}_{\mu}\widehat U^\dag-\frac{i}{g_4}(\partial_\mu \widehat U)\widehat U^\dag,   \lsp \mathbb{U}\to \widehat U e^{ig_4 \widehat \Pi^{(+)}}, \label{gauge_trans_sm}
\eeq
whereas the ${\cal H}^{(\pm)}_0$ are gauge invariant under the 4D residual gauge transformation $\widehat U$.
Equation~\eqref{eff_action_zero_0} is a non-Abelian analog of the Abelian zero-mode
action given by \eqref{eff_action_Abelian_zm}.

We introduce a convenient notion for our effective theory by redefining $V_{0\mu}^{(+)}(x)\equiv V_\mu(x)$, $\pi^{(+)}_{\tilde V0}(x)\equiv \pi_{\tilde V}(x)$, $\widehat \Pi^{(+)}_{0}(x)\equiv \widehat \Pi(x)$ and
\beq
H_1(x)\equiv e^{ig_4 \widehat \Pi(x)}{\cal H}^{(+)}_0(x), \lsp H_2(x)\equiv e^{ig_4 \widehat \Pi(x)}{\cal H}^{(-)}_0(x).  \label{H_12}
\eeq
Now the above effective action \eqref{eff_action_zero_0}, including the  scalar interaction terms, can be written in a nice gauge covariant form as\footnote{Note that the action of Eq.~\eqref{eff_action_nab_gauge_cov} is a non-Abelian version of the
Abelian zero-mode action \eqref{eff_action_ab_gauge_cov}.
}
\begin{align}
S_{eff}=-\int d^4x &\bigg\{\frac14 {\cal F}_{\mu\nu}{\cal F}^{\mu\nu}+\frac14 {\cal Z}_{\mu\nu}{\cal Z}^{\mu\nu}+\frac12 {\cal W}^{+}_{\mu\nu}{\cal W}^{-\mu\nu} \notag\\
&+ \big({\cal D}_\mu H_1\big)^{\dag}{\cal D}^\mu H_1+ \big({\cal D}_\mu H_2\big)^{\dag}{\cal D}^\mu H_2 +V(H_1,H_2) \bigg\},    \label{eff_action_nab_gauge_cov}
\end{align}
where the scalar potential can be written as
\begin{align}
V(H_1,H_2) =&-\mu^{2}|H_1|^2 -\mu^{2}|H_2|^2 +\lambda|H_1|^4 +\lambda|H_2|^4+6\lambda|H_1|^2|H_2|^2.    \label{potenial_sm}
\end{align}
The covariant derivative ${\cal D}_\mu$ is defined as
\begin{align}
{\cal D}_\mu&=\partial_\mu-ig_4\mathbb{\hat A}_{\mu},   \label{co_dir_4_sm}
\end{align}
where $\mathbb{\hat A}_{\mu}$ is defined in Eq.~\eqref{A_matrix_0}.
In the above scalar potential the mass parameter $\mu$ and quartic coupling $\lambda$ are defined as (see Appendix \ref{SSB in the IR-UV-IR model: the Abelian Higgs mechanism}),
\beq
\mu^2\equiv-m_{0}^{2(\pm)}=(1+\beta)m_{KK}^2\delta_{IR}, \Lsp \lambda\equiv\lambda_{IR}(1+\beta)^2, 	\label{mu_lambda}
\eeq
where $\delta_{IR}$, $m_{KK}$ and $\beta$ are defined in Eq.~\eqref{delta_IR_mKK}.

Concerning the symmetries of the above potential, one can notice that $V(H_1,H_2)$ is invariant under $[SU(2)\times U(1)_Y]^\prime \times [SU(2)\times U(1)_Y]$, where one of the blocks has been gauged while the other one survived as a global symmetry.
The zero-modes of the four odd vector bosons $(W_{0\mu}^{(-)\pm}, Z_{0\mu}^{(-)} \text{ and } A_{0\mu}^{(-)})$ and the three would-be-Goldstone
bosons $\Pi^{(-)}_0$ have been removed by appropriate b.c., implying that the
corresponding gauge symmetry has been broken explicitly. What remains is {\it the
truncated inert doublet model}, that contains ${H}_{1,2}$, and the corresponding residual $SU(2)\times U(1)_Y$ global symmetry of the action.
Symmetry under the above mentioned $U(1)^\prime \times U(1)$ implies in particular that
$V(H_1,H_2)$ is also invariant under various $\mathbb{Z}_2$'s, for example
$H_1\to -H_1$, $H_2\to -H_2$ and $H_1\to \pm H_2$.

As explained in the Abelian case~\ref{EWSB by vacuum expectation value of KK modes}, we choose the vacuum such that the even parity Higgs field $H_1$ acquires a vev,
whereas the odd parity Higgs field $H_2$ does not,    i.e.
\beq
v^2_1\equiv v^2=\frac{\mu^2}{\lambda},  \Lsp v_2=0.        \label{v1_v2}
\eeq
Let us now consider fluctuations around the vacuum of our choice
\beq
H_1(x)=\frac{1}{\sqrt2}e^{ig_4\widehat \Pi}\bpm 0\\v+h\epm,    \Lsp H_2(x)=\frac{1}{\sqrt2}e^{ig_4\widehat \Pi}\bpm 0\\ \chi \epm,    \label{H1_H2_def_sm}
\eeq
where $\widehat \Pi$ (defined in Eq. \eqref{pi_matrix_0}) contains the pseudoscalar Goldstone
bosons $\pi_{W^\pm,Z}$. We choose the unitary gauge in which $\pi_{W^\pm,Z}$ are gauged away,    that is they are
eaten up by the massive gauge bosons $W^\pm_\mu$ and $Z_\mu$.
Hence in the unitary gauge our effective action up to the quadratic order
in fluctuations is
\begin{align}
S^{(2)}_{eff}=-\int d^4x &\bigg\{\frac12 {\cal W}^{+}_{\mu\nu}{\cal W}^{-\mu\nu}+ \frac14 {\cal Z}_{\mu\nu}{\cal Z}^{\mu\nu}+\frac14 {\cal F}_{\mu\nu}{\cal F}^{\mu\nu}+ m^2_W W^+_\mu W^{-\mu}+\frac12 m^2_Z Z_\mu Z^\mu\notag\\
&+\frac12\partial_\mu h\partial^\mu h+ \frac12m^2_h h^2  +\frac12\partial_\mu \chi\partial^\mu \chi+ \frac12m^2_\chi \chi^2\bigg\},    \label{eff_action_quadratic}
\end{align}
where the masses are,
\begin{align}
m^2_h&=m^2_\chi=2\mu^2, \lsp m^2_{W}=\frac{1}{4}g^2_4\frac{\mu^2}{\lambda},   \lsp m^2_{Z}=\frac14\Big(g^2_4+g^{\p2}_4\Big)\frac{\mu^2}{\lambda}=\frac{m^2_{W}}{\cos^2\theta_W}.         \label{masses_higgs_WZ}
\end{align}
It is worth noticing here that the Higgs mass $m_h$ and the dark scalar mass $m_\chi$ are degenerate at the tree level.  However, as we demonstrate below, this degeneracy is lifted by the quantum corrections predicted by the effective theory below the KK-mass scale.
The interaction terms for effective theory can be written as
\begin{align}
S_{int}&=-\int d^4x \bigg\{\lambda vh^3+\frac\lambda4 h^4+\frac\lambda4 \chi^4+3\lambda v h\chi^2+\frac32\lambda h^2\chi^2+\frac{g_4^2}{2}vW_\mu^+W^{-\mu}h   \notag\\
&+\frac{g_4^2}{4}W_\mu^+W^{-\mu}(h^2+\chi^2)+\frac14(g_4^2+g_4^{\p2})vhZ_\mu Z^\mu  +\frac18(g_4^2+g_4^{\p2})Z_\mu Z^\mu (h^2+\chi^2)\bigg\}, \label{eff_action_int}
\end{align}
where we have omitted terms involving gauge interactions alone as they are irrelevant to our discussion below.

\subsection{Quantum corrections to scalar masses}
\label{Quantum corrections to scalar masses}
\noindent In this subsection we will study the quantum corrections to the tree-level scalar masses of the Higgs boson
$h$ and the dark matter candidate $\chi$.

Before proceeding further, we want to point out here that in this work we have not studied fermions in our geometric
setup since our focus is on the bosonic sector of the SM and EWSB.
For the sake of self-consistency, we mention here three possibilities for fermion localization and their implications in our geometric setup:
\ben\itemsep0em
\item In this first scenario, one takes the heavy (top) quarks to be localized towards the IR-brane, while the light quarks and leptons are localized towards the UV-brane.
Through this geometric localization one can address the fermion mass hierarchy problem.
In this scenario the even and odd zero-modes corresponding to the heavy quarks
will be almost degenerate in our symmetric geometry, whereas the odd zero-modes corresponding to the light quarks could be
much heavier than their corresponding even zero-modes  \cite{Agashe:2007jb,Medina:2010mu}.
\item In the second scenario, {\it all} the fermions have flat zero-mode profiles. This can be achieved by the choice of appropriate bulk mass parameters for the fermions. As a consequence of flat profiles the odd fermion zero-modes have to disappear
and the even zero-modes will correspond to the SM fermions (in this case the fermion mass hierarchy problem is reintroduced).
\item In the third scenario  {\it all} the fermions are localized towards UV-brane. In this case the masses of {\it all} odd zero-modes of the fermions could be heavier  than their corresponding even zero-modes.
\een
In this study we implicitly limit ourselves to the last two cases in order that the dark Higgs be the lightest odd particle and all the other odd zero-modes
are either not present in our low-energy effective theory or they are much heavier that the dark Higgs, which will therefore be the only relevant dark matter candidate. For either of the choices 2. or 3. above,  the top Yukawa coupling $y_t$ in the low-energy effective
theory will be the same
as in the SM and the top-quark loop correction to the SM Higgs boson mass will be exactly as in the SM up to the KK cutoff.  In case 2., the $n\neq 0$ fermion KK-modes are all much heavier than the KK cutoff, $m_{KK}$,  and will not significantly influence the radiative corrections to the  SM Higgs mass.
We leave the study of the complete fermionic sector associated with our geometric setup for future studies.

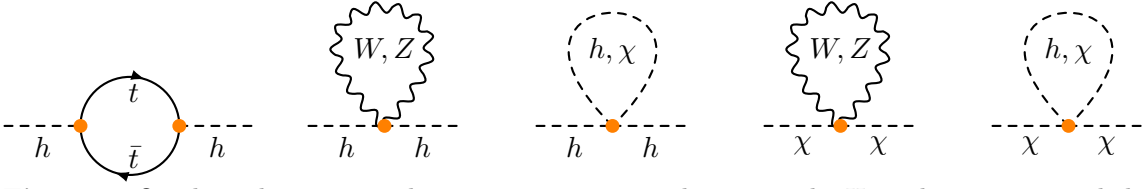
\begin{figure}
\centering
\begin{tikzpicture}[node distance=1cm,very thick, rounded corners=0pt,line cap=round]
\begin{scope}
\coordinate[] (a1);
\coordinate[right=of a1] (v1);
\coordinate[right=1.3cm of v1] (v2);
\coordinate[right=of v2] (a2);
\draw[scalar] (a1) --node[below]{$h$} (v1);
\draw[scalar] (v2)--node[below]{$h$}(a2);
\draw[fermion](v1) arc (180:0:0.65);
\draw[fermion](v2) arc (0:-180:0.65)
(1.7,0.7) node[below]{$t$}
(1.7,-0.7) node[above]{$\bar t$};
\filldraw [orange] (v1) circle (2pt)
                (v2) circle (2pt);
\end{scope}
\begin{scope}[xshift=4cm]
\coordinate[] (a1);
\coordinate[right=of a1] (v1);
\coordinate[right=of v1] (a2);
\draw[scalar] (a1) --node[below]{$h$}(v1);
\draw[scalar] (v1)--node[below]{$h$}(a2);
\draw[boson] (v1) .. controls (-1,2) and (3.3,2) .. (1,-0.2)
(1,1) node[]{$W,Z$};
\filldraw [orange] (v1) circle (2pt);\end{scope}
\begin{scope}[xshift=7cm]
\coordinate[] (a1);
\coordinate[right=of a1] (v1);
\coordinate[right=of v1] (a2);
\draw[scalar] (a1) --node[below]{$h$}(v1);
\draw[scalar] (v1)--node[below]{$h$}(a2);
\draw[scalar] (v1) .. controls (-1,2) and (3,2) .. (v1)
(1,1) node[]{$h,\chi$};
\filldraw [orange] (v1) circle (2pt);
\end{scope}
\begin{scope}[xshift=10cm]
\coordinate[] (a1);
\coordinate[right=of a1] (v1);
\coordinate[right=of v1] (a2);
\draw[scalar] (a1) --node[below]{$\chi$}(v1);
\draw[scalar] (v1)--node[below]{$\chi$}(a2);
\draw[boson] (v1) .. controls (-1,2) and (3.3,2) .. (1,-0.2)
(1,1) node[]{$W,Z$};
\filldraw [orange] (v1) circle (2pt);\end{scope}
\begin{scope}[xshift=13cm]
\coordinate[] (a1);
\coordinate[right=of a1] (v1);
\coordinate[right=of v1] (a2);
\draw[scalar] (a1) --node[below]{$\chi$}(v1);
\draw[scalar] (v1)--node[below]{$\chi$}(a2);
\draw[scalar] (v1) .. controls (-1,2) and (3,2) .. (v1)
(1,1) node[]{$h,\chi$};
\filldraw [orange] (v1) circle (2pt);
\end{scope}
\end{tikzpicture}
\caption{One-loop diagrams in the unitary gauge contributing to the Higgs boson mass and the DM scalar mass.}
\label{loop_diagrams}
\end{figure}

The quantum corrections to the Higgs boson ($h$) mass and the dark-Higgs  ($\chi$) mass within our effective
theory below the KK-scale are quite essential for breaking the mass degeneracy of  Eq.~\eqref{masses_higgs_WZ}. For instance, at the 1-loop level of the perturbative expansion, the main contributions (quadratically divergent) to the masses of the SM
Higgs and the dark-Higgs  come from the exchanges of the top quark ($t$), massive gauge bosons
 ($W,~Z$), Higgs boson ($h$) and the dark-Higgs ($\chi$) in the loop~\footnote{Another scalar which could be potentially present in our effective theory is the {\it radion}, which is responsible for the stabilization of the setup. The stabilization mechanism is beyond the scope of the present work, as here we assume a rigid geometrical background with no fluctuations of the 5D metric. However, we want to comment here that if the radion were present in our effective theory, because of it bosonic nature it would likely reduce the fine-tuning much in the manner that the $\chi$ does.}, see Fig. \ref{loop_diagrams}.
It is instructive to write the general 1-loop effective scalar potential $V_{eff}(H_1,H_2)$ for our effective theory, described in the previous section, as~\footnote{Note that in this section we are considering the Higgs doublets $H_{1,2}$ in the unitary gauge, such that $H_1(x)=\frac{1}{\sqrt2}\bpm 0\\v_1+h\epm$ and $H_2(x)=\frac{1}{\sqrt2}\bpm 0\\ v_2+\chi \epm$, where at tree level our choice was $v_1=v$ and $v_2=0$.}
\beq
V_{eff}(H_1,H_2)=V_0(H_1,H_2)+V_1(H_1,H_2),       \label{effective_potential}
\eeq
where $V_0(H_1,H_2)$ is the tree level scalar potential given by Eq. \eqref{potenial_sm} and $V_1(H_1,H_2)$ is the 1-loop effective potential, given by (see for example Refs. \cite{Grzadkowski:2009mj,Kolda:2000wi,Casas:2004gh})
\begin{align}
V_1(H_1,H_2)&=\frac{\Lambda^2}{32\pi^2}\left[ 3\Big(g_4^2+\frac12(g_4^2+g^{\p2}_4)+8\lambda\Big)(|H_1|^2+|H_2|^2)-12y_t^2|H_1|^2\right] + \cdots,   \label{potential_loop}
\end{align}
where $y_t$ is the top Yukawa coupling, related to top mass through $m_t^2=y_t^2v^2/2$.
We use the momentum cut-off regularization. Also it is important to comment here
that $H_2$ is odd under the geometric $\mathbb{Z}_2$ parity, implying that it does not couple
to the even zero-mode fermions. Moreover, we consider only the quadratically divergent part
of the effective scalar potential and the ellipses in the above equation
represent the terms which are not quadratically divergent.

\begin{figure}
\begin{center}
\includegraphics[scale=0.45]{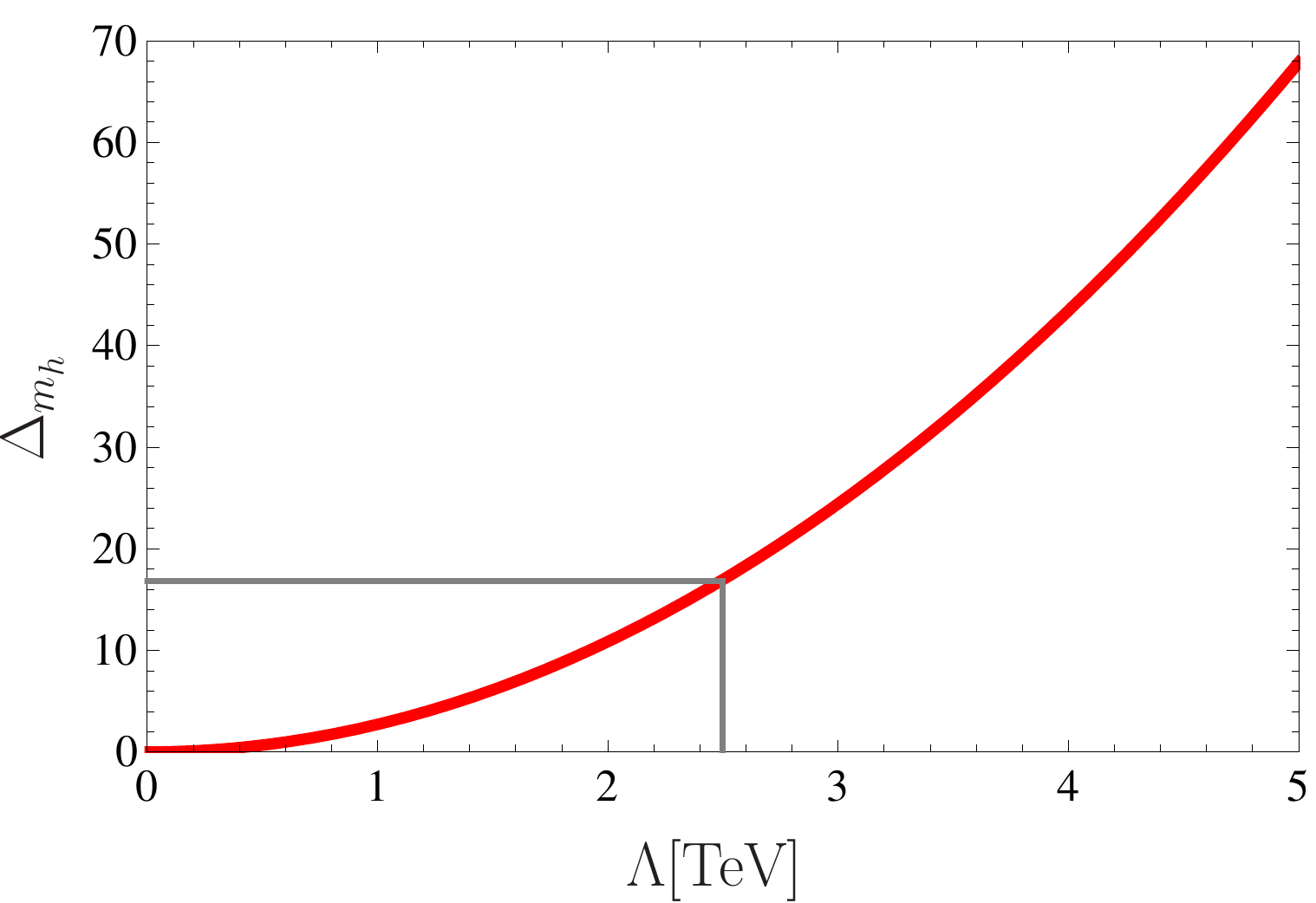}\hsp
\includegraphics[scale=0.45]{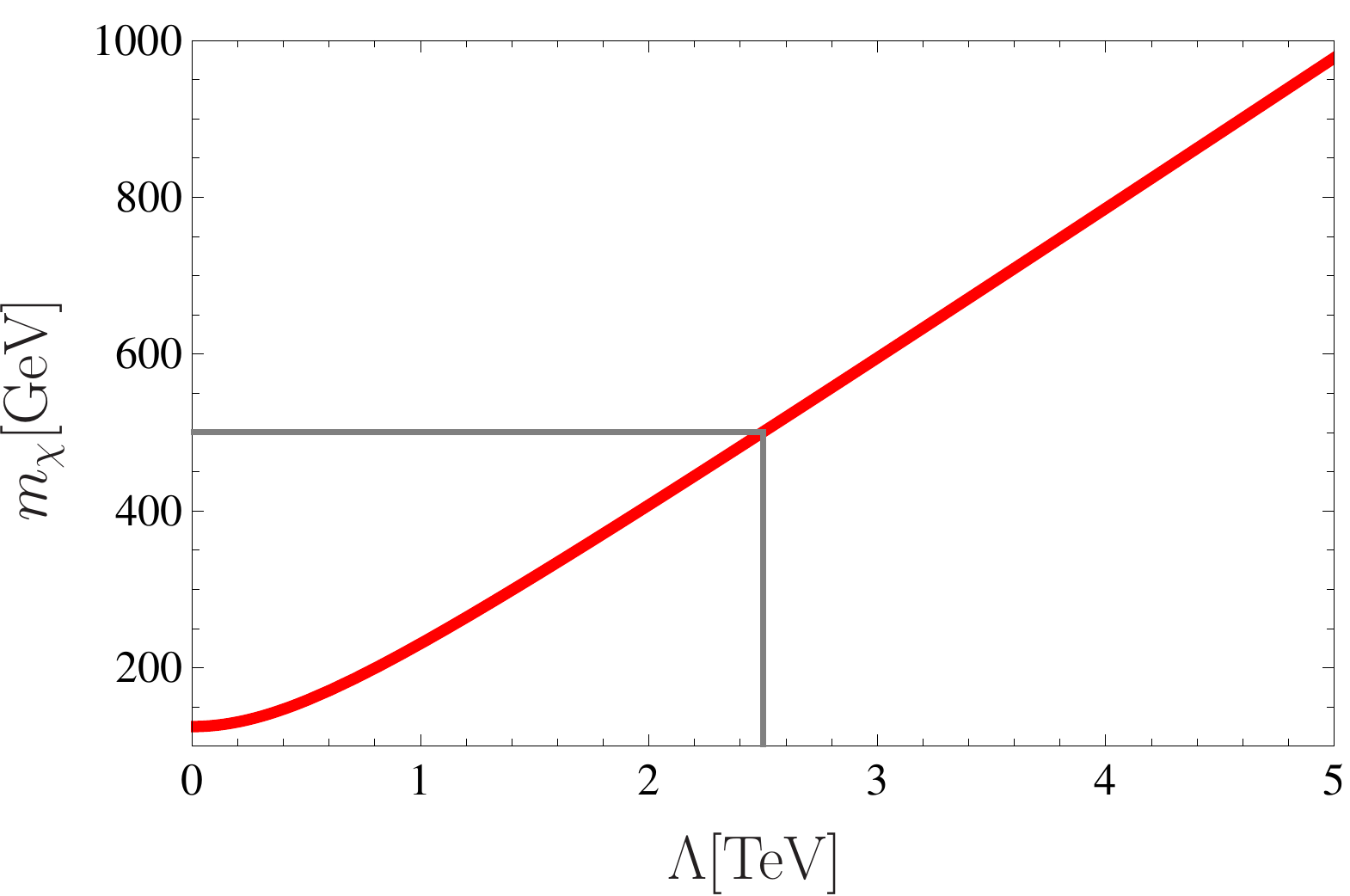}
\end{center}
\caption{The left plot gives the value of the fine-tuning measure $\Delta_{m_h}$ for a Higgs mass of $125$ GeV as a function of the cutoff $\Lambda$.  The  right plot shows the dependence of $m_\chi$ on $\Lambda$ for $m_h=125\gev$. In our model $\Lambda=m_{KK}$.  The vertical gray line indicates the  current lower bound on the KK mass scale coming from  EWPT as computed in our model for $\beta=0$,    $m_{KK}\gtrsim2.5\tev$.}
\label{mchi_lambda}
\end{figure}

The minimization of the effective potential $V_{eff}(H_1,H_2)$,    i.e.
\begin{equation}
\frac{\partial V_{eff}}{\partial H_i}\Big\vert_{H_i=\langle H_i\rangle}=0,  \hsp\text{where}\hsp \langle H_i\rangle=\frac{1}{\sqrt2}\bpm0\\v_{i}\epm \hsp i=1,2
\end{equation}
gives the following set of conditions for the global minimum,
\beq
\lambda v_1^2=\mu^2-\delta\mu^2-3\lambda v_2^2, \hsp \text{or}\hsp v_1=0,       \label{v1_eff}\\
\eeq
and
\beq
\lambda v_2^2=\mu^2-\delta\mu^2 + \frac38\frac{\Lambda^2}{\pi^2}y_t^2 -
3\lambda v_1^2, \hsp \text{or}\hsp v_2=0,       \label{v2_eff}
\eeq
where $\delta\mu^2$ is given by
\beq
\delta\mu^2=\frac{3\Lambda^2}{32\pi^2}\Big[g_4^2+\frac12(g_4^2+g^{\p2}_4)+8\lambda-4y_t^2\Big].  \label{delta_mu}
\eeq
Of the four possible global minima of Eqs.~\eqref{v1_eff} and \eqref{v2_eff}, we will choose the vacuum such that  $H_1$ acquires the vev, whereas $H_2$ does not:
\[
v_1=v, \Lsp v_2=0,
\]
where $v\simeq246 \gev$ is the vacuum expectation value of the SM Higgs doublet.
With this choice of vacuum,  the 1-loop corrected masses for the fluctuations around the vacuum are
\begin{align}
m^2_h=\frac{\partial^2V_{eff}(H_1,H_2)}{\partial H_1^2}\Big\vert_{H_1=v,H_2=0}&=\Big(-\mu^2+\delta\mu^2\Big)+3\lambda v^2=2\lambda v^2,    \label{hat_mh}\\
m^2_\chi=\frac{\partial^2V_{eff}(H_1,H_2)}{\partial H_2^2}\Big\vert_{H_1=v,H_2=0}&=\Big(-\mu^2+\delta\mu^2\Big)+3\lambda v^2+\frac38\frac{\Lambda^2}{\pi^2}y_t^2,     \notag\\
&=2\lambda v^2+\frac34\frac{\Lambda^2}{\pi^2v^2}m_t^2.    \label{hat_mchi}
\end{align}
To get $m_h=125\gev$, equivalent to $v\simeq 246\gev$, we need to fine-tune the parameters of the theory. To quantify the level of fine-tuning, we employ the  Barbieri--Giudice fine-tuning measure $\Delta_{m_h}$ \cite{Barbieri:1987fn,Kolda:2000wi,Casas:2004gh}:
\beq
\Delta_{m_h}\equiv \left\vert\frac{\delta\mu^2}{\mu^2}\right\vert=\left\vert\frac{\delta m_h^2}{m_h^2}\right\vert.  \label{fine-tuning}
\eeq
We plot the fine-tuning measure $\Delta_{m_h}$ as a function of the effective cutoff scale $\Lambda$ in Fig. \ref{mchi_lambda}. If one allows fine-tuning of about 10\%,    i.e.  $\Delta_{m_h}=10$, then the effective cutoff scale is $\Lambda\simeq2\tev$. The most stringent bounds on the KK-scale $m_{KK}$ in RS1 geometry with a bulk Higgs come from  electroweak precision tests (EWPT) by fitting the $S,~T$ and $U$ parameters \cite{Archer:2014jca}. The lower bound on the KK mass scale in our model (AdS geometry,    i.e.  $A(y)=-k|y|$) is $m_{KK}\gtrsim2.5\tev$ for $\beta=0$ and $m_{KK}\gtrsim4.3\tev$ for $\beta=10$ at $95\%$ C.L. \cite{Archer:2014jca}. This implies a tension between fine-tuning (naturalness) and the lower bound on the KK mass scale $m_{KK}$. The region within the gray lines in Fig. \ref{mchi_lambda} shows the current bounds on the KK mass scale for our geometry and the associated fine-tuning. It is important to comment here that a slight modification to the AdS geometry (for example, models with soft wall or thick branes) leads to a considerable relaxation of the above mentioned lower bound on KK mass scale \cite{Cabrer:2010si,Cabrer:2011fb,Cabrer:2011vu}. For instance, a mild modification to the AdS metric in the vicinity of the IR-brane can relax the KK mass scale to $m_{KK}\gtrsim 1\tev$ \cite{Cabrer:2010si,Cabrer:2011fb,Cabrer:2011vu,Iyer:2015ywa,Carmona:2011ib}. Needless to say, the generalization of our model to modified AdS geometries with soft walls or thick branes is possible.

The 1-loop quantum corrected dark matter squared mass $m^2_\chi$ is:
\beq
m_\chi^2=m_h^2+ \frac34\frac{\Lambda^2}{\pi^2v^2}m_t^2\,.
\label{m-shift}
\eeq
Hence,  $m_\chi$ is raised linearly with the cut-off scale $\Lambda$. This is illustrated in Fig.~\ref{mchi_lambda}. An interesting aspect of our model is that dark matter is predicted to be heavier than the SM Higgs boson.
A natural value of the cutoff coincides with the mass of the first KK excitations, which are experimentally
limited~\cite{Agashe:2014kda}  to lie above a few TeV (depending on model details and KK mode sought).  Requiring that the fine-tuning measure $\Delta_{m_h}$ be less than $100$ implies that $m_{KK}$ should be below about 6 TeV. Meanwhile, the strongest version of the EWPT bound requires $m_{KK}\gsim 2.5\tev$, corresponding to $m_\chi\sim 500\gev$, for which $\Delta_{m_h}$ is a very modest $\sim 18$. In short, our model is most consistent for $500\gev\lsim m_\chi \lsim 1200\gev$.

\subsection{Dark matter relic abundance}
\label{Dark matter relic abundance}
\begin{figure}
\centering
\begin{tikzpicture}[node distance=1cm,very thick, rounded corners=0pt,line cap=round]
\begin{scope}[xshift=0.5cm]
\coordinate[] (v1);
\coordinate[above left=of v1] (a1);
\coordinate[below left=of v1] (a2);
\coordinate[above right=of v1] (b1);
\coordinate[below right=of v1] (b2);
\draw[scalar] (a1)node[left]{$\chi$} -- (v1);
\draw[scalar] (a2)node[left]{$\chi$} -- (v1);
\draw[boson](v1)--(b1)node[right]{$W,Z$};
\draw[boson](v1)--(b2)node[right]{$W,Z$};
\filldraw [orange] (v1) circle (2pt);
\end{scope}
\begin{scope}[xshift=4.5cm]
\coordinate[] (v1);
\coordinate[above left=of v1] (a1);
\coordinate[below left=of v1] (a2);
\coordinate[right=1.3cm of v1] (v2);
\coordinate[above right=of v2] (b1);
\coordinate[below right=of v2] (b2);
\draw[scalar] (a1)node[left]{$\chi$} -- (v1);
\draw[scalar] (a2)node[left]{$\chi$} -- (v1);
\draw[scalar] (v1)--node[above]{$h$}(v2);
\draw[boson](v2)--(b1)node[right]{$W,Z$};
\draw[boson](v2)--(b2)node[right]{$W,Z$};
\filldraw [orange] (v1) circle (2pt)
                (v2) circle (2pt);
\end{scope}
\begin{scope}[xshift=9.5cm]
\coordinate[] (v1);
\coordinate[above left=of v1] (a1);
\coordinate[below left=of v1] (a2);
\coordinate[right=1.3cm of v1] (v2);
\coordinate[above right=of v2] (b1);
\coordinate[below right=of v2] (b2);
\draw[scalar] (a1)node[left]{$\chi$} -- (v1);
\draw[scalar] (a2)node[left]{$\chi$} -- (v1);
\draw[scalar] (v1)--node[above]{$h$}(v2);
\draw[fermion](v2)--(b1)node[right]{$f$};
\draw[fermion](b2)node[right]{$\bar f$} -- (v2);
\filldraw [orange] (v1) circle (2pt)
                (v2) circle (2pt);
\end{scope}\newline
\begin{scope}[yshift=-2.5cm]
\coordinate[] (v1);
\coordinate[above left=of v1] (a1);
\coordinate[below left=of v1] (a2);
\coordinate[above right=of v1] (b1);
\coordinate[below right=of v1] (b2);
\draw[scalar] (a1)node[left]{$\chi$} -- (v1);
\draw[scalar] (a2)node[left]{$\chi$} -- (v1);
\draw[scalar](v1)--(b1)node[right]{$h$};
\draw[scalar](v1)--(b2)node[right]{$h$};
\filldraw [orange] (v1) circle (2pt);
\end{scope}
\begin{scope}[xshift=3cm,yshift=-2.5cm]
\coordinate[] (v1);
\coordinate[above left=of v1] (a1);
\coordinate[below left=of v1] (a2);
\coordinate[right=1.3cm of v1] (v2);
\coordinate[above right=of v2] (b1);
\coordinate[below right=of v2] (b2);
\draw[scalar] (a1)node[left]{$\chi$} -- (v1);
\draw[scalar] (a2)node[left]{$\chi$} -- (v1);
\draw[scalar] (v1)--node[above]{$h$}(v2);
\draw[scalar](v2)--(b1)node[right]{$h$};
\draw[scalar](v2) -- (b2)node[right]{$h$};
\filldraw [orange] (v1) circle (2pt)
                (v2) circle (2pt);
\end{scope}
\begin{scope}[xshift=7.5cm,yshift=-2cm]
\coordinate[] (v1);
\coordinate[below= of v1] (v2);
\coordinate[position=150 degrees from v1] (a1);
\coordinate[position=-150 degrees from v2] (a2);
\coordinate[position=30 degrees from v1] (b1);
\coordinate[position=-30 degrees from v2] (b2);
\draw[scalar] (a1)node[left]{$\chi$} -- (v1);
\draw[scalar] (a2)node[left]{$\chi$} -- (v2);
\draw[scalar] (v1)--node[left]{$\chi$}(v2);
\draw[scalar](v1)--(b1)node[right]{$h$};
\draw[scalar](v2)--(b2)node[right]{$h$};
\filldraw [orange] (v1) circle (2pt)
                (v2) circle (2pt);
\end{scope}
\begin{scope}[xshift=11cm,yshift=-2cm]
\coordinate[] (v1);
\coordinate[below= of v1] (v2);
\coordinate[position=150 degrees from v1] (a1);
\coordinate[position=-150 degrees from v2] (a2);
\coordinate[above right=1.8cm of v2] (b1);
\coordinate[below right=1.8cm of v1] (b2);
\draw[scalar] (a1)node[left]{$\chi$} -- (v1);
\draw[scalar] (a2)node[left]{$\chi$} -- (v2);
\draw[scalar] (v1)--node[left]{$\chi$}(v2);
\draw[scalar](v2)--(b1)node[right]{$h$};
\draw[scalar](v1)--(b2)node[right]{$h$};
\filldraw [orange] (v1) circle (2pt)
                (v2) circle (2pt);
\end{scope}
\end{tikzpicture}
\caption{Dark matter annihilation diagrams.}
\label{DM_annihi_diagrams}
\end{figure}
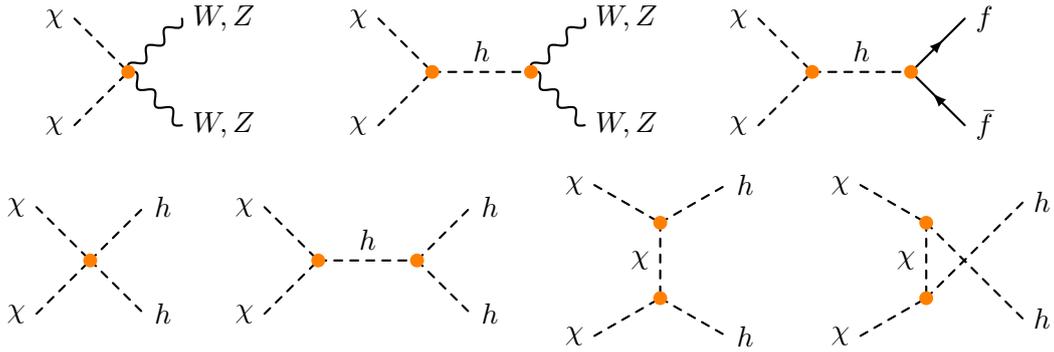

\noindent In this subsection we calculate the dark matter relic abundance. The diagrams contributing to dark matter annihilation are shown in Fig.~\ref{DM_annihi_diagrams}. The squared amplitudes $|{\cal M}|^2$ corresponding to the contribution of each final state  to dark matter annihilation are:
\begin{align}
\left| \mathcal{M} (\chi\chi \to \tilde V\tilde V)\right|^2 &= {4 m_{\tilde V}^4 \over S_{\tilde V} v^4} \left( 1+ {3m_h^2 \over s-m_h^2} \right)^2 \left[ 2+ \left( 1- {s \over 2m^2_{\tilde V}} \right)^2 \right],  \label{M2VV}\\
\left| \mathcal{M} (\chi\chi \to f\bar{f}) \right| ^2&=  18 N_c {m_f^2 m_h^4 \over v^4} {s-4m^2_f \over (s-m_h^2)^2}, \label{M2ff}\\
\left| \mathcal{M} (\chi\chi \to hh) \right|^2 &= {9 m_h^4 \over 2 v^4} \left[ 1+ 3m_h^2 \left( {1 \over s-m_h^2} + {1 \over t-m_\chi^2} + {1 \over u-m_\chi^2} \right) \right]^2,  \label{M2hh}
\end{align}
where $\tilde V={W,Z}$ and $S_W=1$ and $S_Z=2$ are the symmetry factors accounting for the identical particles in the final state; $N_c$ refers to the number of ``color'' degrees of freedom for the given fermion and $s,~t,~u$ are the Mandelstam variables. Here, we ignore the loop-induced $\gamma\gamma$ and $Z\gamma$ final states, which are strongly suppressed.
Note that the first term in the parenthesis in Eq.~(\ref{M2VV}) and the first term in the square bracket in Eq.~(\ref{M2hh}) arise from the $\chi\chi \tilde V\tilde V$ and the $\chi\chi hh$ contact interactions, respectively.
The former channel is present in our model since $\chi$ is a component of the  (truncated) odd $SU(2)$ doublet.

\begin{figure}
\begin{center}
\includegraphics[width=0.47\textwidth]{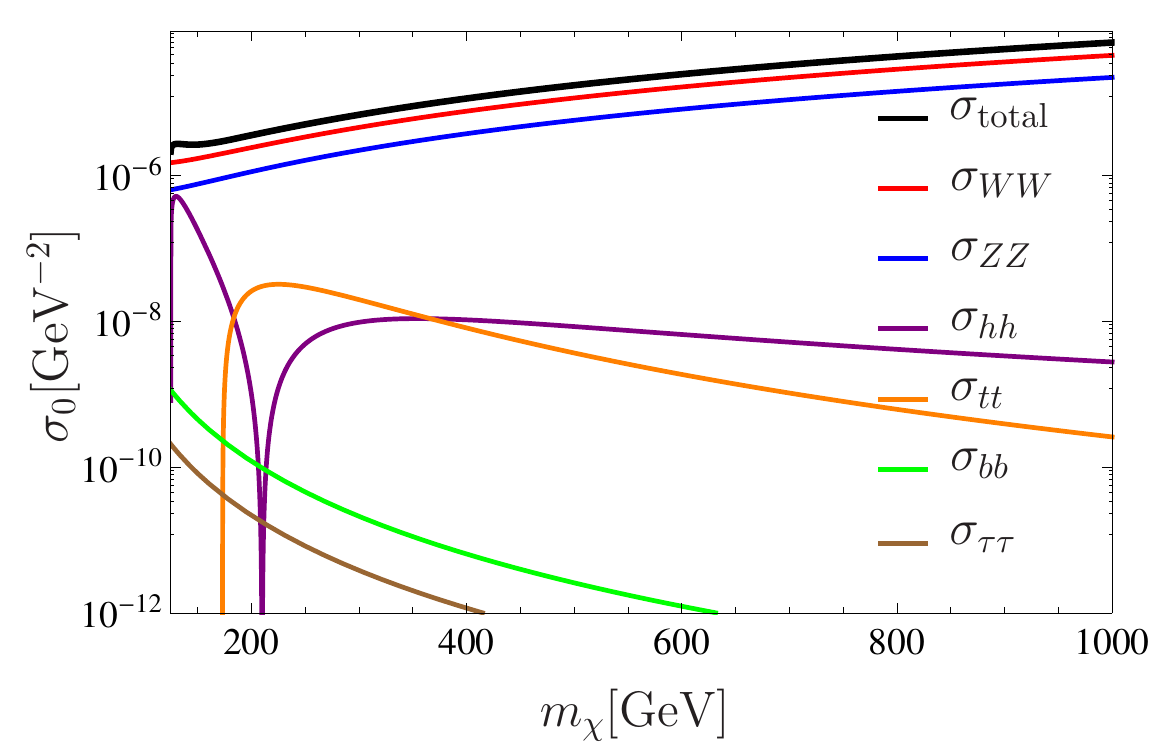}\hsp
\includegraphics[width=0.45\textwidth]{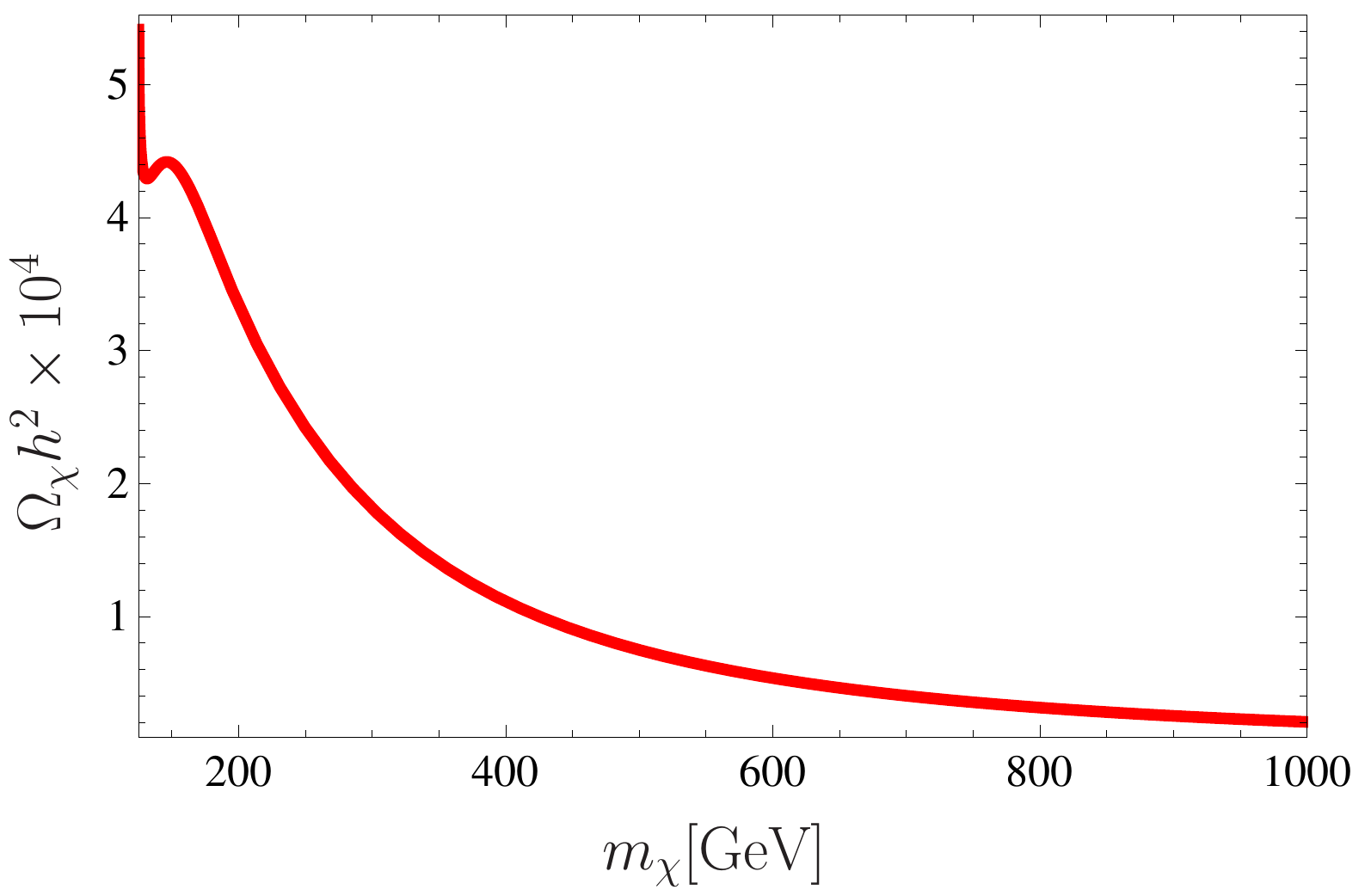}
\end{center}
\caption{The above graphs show the annihilation cross-section $\sigma_0$ for different final states (left)
and the $\chi$ abundance $\Omega_\chi h^2\times 10^{4}$ (right) as a function of dark matter mass $m_\chi$.}
\label{sigma_omega}
\end{figure}

In the left panel of Fig. \ref{sigma_omega} we plot the annihilation cross-section for
the contributing channels as a function of $m_\chi$.  (Note that the parameter $\Lambda$ would only enter if we performed this calculation at the one-loop level.)
As seen from the plot, the total cross section is dominated by $WW$ and $ZZ$ final states.
The main contributions for these final states are those generated by the contact interactions $\chi\chi \tilde V\tilde V$.
In fact, in our model, the $\tilde V\tilde V$ final states are additionally enhanced by a constructive interference of
the contact $\chi\chi \tilde V\tilde V$ interaction with the s-channel Higgs-exchange diagram.
In addition, for low $m_\chi$, there is a comparable contribution from $\chi\chi$ annihilation into $hh$.
(The dip at $m_\chi\sim 210\gev$ is caused by cancellation between the contact
interaction and $s,~t,~u$-channel diagrams.) Fermionic final states are always irrelevant;
even  $\chi\chi\to t \bar t$ production is very small in comparison to $\chi\chi \to \tilde V\tilde V$.
Then, adopting the standard s-wave cold dark matter approximation \cite{Kolb:1990vq}, we
find the present $\chi$ abundance $\Omega_\chi h^2$  shown in the right panel of Fig. \ref{sigma_omega}. We observe that $\Omega_\chi h^2\lsim 10^{-4} $ once the EWPT bound of $m_\chi\gsim 500\gev$ is imposed.
Clearly, some other dark matter component is needed within this model to satisfy the Planck
measurement, $\Omega_\chi h^2 \sim 0.1$ \cite{Planck:2015xua}.

\section{Summary}
\label{Summary}
\noindent In this paper, we constructed a model with ${\mathbb Z}_2$ geometric symmetry such that two identical AdS patches are glued together at $y=0$, where $y$ is the coordinate of the fifth dimension. We considered three D3-branes, one at $y=0$ referred to as the UV-brane where gravity is assumed to be localized and two branes at $y=\pm L$ referred to as IR-branes -- the IR-UV-IR model. For this $\mathbb{Z}_2$ symmetric geometric setup we found that the RS metric \eqref{metric} is the background solution  of pure gravity when the matter backreaction is neglected. To investigate possible backreaction of matter fields (bulk $SU(2)$ Higgs doublet) on the geometry, we solved the 5D coupled scalar-gravity equations of motion adopting the superpotential method. We found analytic solutions where the Higgs background vev is highly localized towards the IR brane. It was also verified that the backreaction is negligible for $kL\simeq 37$ which is required in order to address the gauge hierarchy problem. The technique employed to find solutions  is very general and can be used to any bulk Higgs RS1-like constructions to stabilize geometry taking into account backreaction.

The motivation of this work is twofold: ({\it i}) to analyze the situation where EWSB is due to the bulk Higgs in this $\mathbb{Z}_2$ symmetric geometry; and ({\it ii}) to discuss the lowest odd KK-mode as a dark matter candidate. Concerning EWSB, we discussed in detail many important aspects of SSB due to a bulk Higgs. In the Appendix \ref{SSB in the IR-UV-IR model: the Abelian Higgs mechanism} we considered the Abelian gauge group where we discussed two alternative approaches to  SSB due to a bulk Higgs field acquiring a vacuum expectation value. In one approach, the symmetry breaking is triggered by a vev of the KK zero-mode  of the bulk Higgs field.  The second approach is based on the expansion of the bulk Higgs field around an extra-dimensional vev with non-trivial $y$ profile. The comparison between the two Abelian scenarios is summarized in Tables \ref{comparison} and \ref{comp_para}. The (zero-mode) effective theories obtained from the two approaches are identical and the most intriguing feature of the Abelian Higgs mechanism is that the even and odd Higgs zero-modes have degenerate mass at the tree-level --- a feature that is also present in the SM case.

To achieve SSB,  the choice of boundary conditions for the fields at $\pm L$ is critical.  In both approaches to the Abelian case, we allowed the $y$-derivative of a field to have an arbitrary value at $\pm L$ as opposed to requiring that the field value itself be zero, i.e. we employed Neumann or mixed b.c.  rather than Dirichlet b.c.  at $\pm L$.  The latter choice would have led to an explicit symmetry breaking scenario in which there are no Goldstone modes and the gauge bosons do not acquire mass. (Note that the boundary or ``jump'' conditions at $y=0$ follow from the bulk equations of motion in the case of even modes, whereas odd modes are required to be zero by symmetry.)

Following this introductory material in Appendix \ref{SSB in the IR-UV-IR model: the Abelian Higgs mechanism}, we considered EWSB assuming that the  SM gauge group is present in the bulk of our $\mathbb{Z}_2$ symmetric 5D warped model. The zero-mode effective theory appropriate at scales below the KK scale,  $m_{KK}$, was obtained. For appropriate  Higgs field potentials in the bulk and localized at the UV and IR-branes, SM-like EWSB is obtained when only the IR-branes have a quartic potential term.  In contrast, quadratic  mass-squared terms are allowed both on the branes and in the bulk. Of course, to achieve spontaneous EWSB, we employed the same boundary conditions as delineated above for the Abelian model. The resulting model has the following features.
\bit \itemsep0em
\item Due to warped KK-parity all fields develop even and odd towers of KK-modes in the 4D effective theory.
\item Assuming that the KK-scale is high enough ($m_{KK}\sim\co(\text{few})\tev$), we derive the low energy effective theory which includes only zero-modes of the theory.
\item In the effective theory, the symmetry of the model is $[SU(2)\times U(1)_Y]^\prime \times [SU(2)\times U(1)_Y]$, where the unprimed symmetry group is gauged while the other stays as a global symmetry.
The zero-mode odd gauge fields and the corresponding Goldstone modes from the odd Higgs doublet are {\it eliminated} due to the b.c..
\item In the low energy effective theory, we are left with all the SM fields plus a {\it dark-Higgs} -- the odd zero-mode Higgs. This dark-Higgs and the SM Higgs (the even zero-mode) are degenerate at tree level.
\item In order to get the SM Higgs mass of $125\gev$, we need to fine-tune the 5D fundamental parameters of theory to about $1\% - 5\%$, where the upper bound is determined by the lower bound on the KK scale coming from EWPT requirements.
\item We computed the one-loop quantum corrections to the tree-level masses of the SM Higgs and the dark Higgs assuming that the cutoff scale of our effective theory is the KK-scale, $m_{KK}$. One finds that the dark-Higgs mass is necessarily larger than the SM Higgs mass, the difference being quadratically dependent on $m_{KK}$.
\item Requiring that the fine-tuning measure $\Delta_{m_h}$ be less than $100$ implies that $m_{KK}$ should be below about 6 TeV. Meanwhile, the strongest version of the EWPT bound requires $m_{KK}\gsim 2.5\tev$, corresponding to $m_\chi\sim 500\gev$, for which $\Delta_{m_h}$ is a very modest $\sim 18$. In short, our model is most consistent for $500\gev\lsim m_\chi \lsim 1200\gev$.
\item We calculate the relic abundance of the dark-Higgs in the cold dark matter approximation. For $m_\chi$ in the above preferred range,  $\Omega_\chi h^2 \lsim 10^{-4}$ as compared to the current experimental value of $\sim 0.1$. To obtain a more consistent dark matter density, one needs to either assume another DM particle or perform a more rigorous analysis of our model by considering the even and odd higher KK-modes in the effective theory.
\eit

\section*{Acknowledgements}
\noindent JFG and YJ are supported in part by US DOE grant DE-SC-000999. AA and BG acknowledge partial support by the National Science Centre (Poland) research project, decision no DEC-2014/13/B/ST2/03969.
The work of AA was partially supported by the Foundation for Polish Science International PhD Projects Programme
co-financed by the EU European Regional Development Fund. AA and BG are grateful to the Mainz Institute for Theoretical Physics (MITP) and the University of California Davis for their hospitality and partial support during the completion of this work.
JFG and YJ acknowledge hospitality and partial support by Warsaw University during the course of the project.
AA would like to thank Adam Falkowski, Matthias Neubert and Mariano Quiros for useful discussions.

\appendix

\section{SSB in the IR-UV-IR model: the Abelian Higgs mechanism}
\label{SSB in the IR-UV-IR model: the Abelian Higgs mechanism}
\noindent In this Appendix we will discuss the mechanism of spontaneous symmetry breaking (SSB) for an
Abelian case with the Higgs field (a complex scalar) in the IR-UV-IR geometry of
Sec.~\ref{IR-UV-IR background solution}. The metric is given
by Eq. \eqref{metric}, we will neglect the back reaction of the bulk fields on the geometry.
We start by specifying the 5D Abelian action,
\begin{align}
S_{Ab}=-\int d^5x \sqrt{-g}\bigg\{&\frac14 F_{MN}F^{MN}+\left( D_M \Phi\right)^\ast D^M \Phi +\mu_B^2\Phi^\ast \Phi   \notag\\
&+V_{IR}(\Phi)\delta(y+L)+V_{UV}(\Phi)\delta(y)+V_{IR}(\Phi)\delta(y-L)\bigg\},    \label{action_5d_c}
\end{align}
where $D_M\equiv \partial_M-ig_5A_M$ with the 5D $U(1)$ coupling constant $g_5$ and $F_{MN}\equiv\partial_M A_N-\partial_N A_M$. We require that the bulk potential and the UV-brane potential have only quadratic terms whereas the IR-brane potential is allowed to have a quartic term:
\beq
V_{UV}(\Phi)=\frac{m^2_{UV}}{k}\Phi^\ast \Phi,   \Lsp V_{IR}(\Phi)=-\frac{m^2_{IR}}{k}\Phi^\ast \Phi+\frac{\lambda_{IR}}{k^2}\left(\Phi^\ast \Phi\right)^2.   \label{boudary_potentials_ab}
\eeq
 In this way EWSB is mainly triggered by the IR-brane. Above,
$\Phi$ is a complex scalar field and the parametrization is such that $m_{UV}$ and $m_{IR}$ have mass dimensions while $\lambda_{IR}$ is dimensionless. The gauge transformations can be written as
\begin{align}
\Phi(x,y)&\to \Phi^\p(x,y)=e^{i\Lambda(x,y)}\Phi(x,y),        \label{gauge_trans_H}\\
A_M(x,y)&\to A_M^\p(x,y)=A_M(x,y)+\frac{1}{g_5}\partial_M \Lambda(x,y),        \label{gauge_trans_A}
\end{align}
where $\Lambda(x,y)$ is the gauge parameter.

As discussed in Sec. \ref{KK-Parity}, the fields in the IR-UV-IR setup have even and odd bulk wave functions implied by the geometric KK-parity. Hence, it is convenient to decompose the
generic Higgs and the gauge field into fields of definite parity as follows
\beq
\Phi(x,y)=\Phi^{(+)}(x,y)+\Phi^{(-)}(x,y),      \lsp A_M(x,y)=A^{(+)}_{M}(x,y)+A^{(-)}_{M}(x,y),     \label{even_odd_HA}
\eeq
where $\pm$ denotes the even and odd states. The gauge transformations for the even and odd parity modes are,
\begin{align}
A^{(\pm)}_\mu(x,y)\to A_\mu^{^\p(\pm)}(x,y)=&A^{(\pm)}_\mu(x,y)+\frac{1}{g_5}\partial_\mu \Lambda^{(\pm)}(x,y),        \label{gauge_trans_A_mu}\\
A^{(\pm)}_5(x,y)\to A_5^{^\p(\pm)}(x,y)=&A^{(\pm)}_5(x,y)+\frac{1}{g_5}\partial_5 \Lambda^{(\mp)}(x,y).        \label{gauge_trans_A_5}\\
\bpm \Phi^{(+)}(x,y) \\ \Phi^{(-)}(x,y)\epm \to \bpm \Phi^{^\p(+)} (x,y)\\ \Phi^{^\p(-)}(x,y)\epm =&e^{i\Lambda^{(+)}(x,y)\mathds{1}} e^{i\Lambda^{(-)}(x,y)\tau_1} \bpm \Phi^{(+)}(x,y) \\ \Phi^{(-)}(x,y)\epm,    \label{H_pm_transformation}
\end{align}
where $\mathds{1}$ is a $2\times2$ unit matrix, whereas $\tau_1$ is the first Pauli matrix.

With this decomposition the above action can be written as
\begin{align}
S_{Ab}=-\int d^5x& \sqrt{-g}\bigg\{\frac14 F^{(+)}_{\mu\nu}F_{(+)}^{\mu\nu}+\frac12 F^{(+)}_{\mu5}F_{(+)}^{\mu5}+ D_M \Phi^{(+)\ast}D^M\Phi^{(+)} +\mu_B^2\Phi^{(+)\ast} \Phi^{(+)}   \notag\\
&+\frac14 F^{(-)}_{\mu\nu}F_{(-)}^{\mu\nu}+\frac12 F^{(-)}_{\mu5}F_{(-)}^{\mu5}+ D_M \Phi^{(-)\ast}D^M\Phi^{(-)} +\mu_B^2\Phi^{(-)\ast} \Phi^{(-)}\notag\\
&+V_{IR}(\Phi^{(\pm)})\delta(y+L)+V_{UV}(\Phi^{(+)})\delta(y)+V_{IR}(\Phi^{(\pm)})\delta(y-L)\bigg\},    \label{action_5d_pm}
\end{align}
where,
\beq
F^{(\pm)}_{\mu\nu}\equiv\partial_\mu A^{(\pm)}_\nu-\partial_\mu A^{(\pm)}_\nu,   \lsp    F^{(\pm)}_{\mu5}\equiv\partial_\mu A^{(\pm)}_5-\partial_5 A^{(\mp)}_\mu.
\eeq
The brane localized potentials for the Higgs field,  $V_{UV}(\Phi)$ and $V_{IR}(\Phi)$, can be written in terms of even and odd parity modes
as
\begin{align}
V_{UV}(\Phi^{(+)})=&\frac{m^2_{UV}}{k}|\Phi^{(+)}|^2,    \label{bp_ab_UV}\\
V_{IR}(\Phi^{(\pm)})=&-\frac{m^2_{IR}}{k}|\Phi^{(+)}|^2-\frac{m^2_{IR}}{k}|\Phi^{(-)}|^2+\frac{\lambda_{IR}}{k^2}|\Phi^{(+)}|^4+\frac{\lambda_{IR}}{k^2}|\Phi^{(-)}|^4 \notag\\
& +\frac{4\lambda_{IR}}{k^2}|\Phi^{(+)}|^2|\Phi^{(-)}|^2+\frac{\lambda_{IR}}{k^2}\left((\Phi^{(+)\ast} \Phi^{(-)})^2+h.c.\right) \,.   \label{bp_ab_IR}
\end{align}
Above, we have not written $\Phi^{(-)}$ terms in  $V_{UV}$  since  $\Phi^{(-)}(0)=0$. Moreover, we have not written terms in the above action, including the potentials,
which are explicitly odd as they will not contribute after integration over the $y$-coordinate. One can
easily check that the above brane potentials are invariant under the gauge transformations defined above.
Also note that  $F^{(\pm)}_{\mu\nu}$ and $F^{(\pm)}_{\mu5}$ are gauge invariant under the gauge
transformations \eqref{gauge_trans_A_mu} and \eqref{gauge_trans_A_5}.
In the even/odd basis, the covariant derivatives $D_\mu$ and $D_5$ following from $D_M\equiv \partial_M-ig_5A_M$, take the form
\begin{align}
D_\mu \bpm \Phi^{(+)} \\ \Phi^{(-)}\epm&\equiv \left[\partial_\mu -ig_5\bpm A^{(+)}_\mu & A^{(-)}_\mu \\
A^{(-)}_\mu &A^{(+)}_\mu \epm \right] \bpm \Phi^{(+)} \\ \Phi^{(-)}\epm,   \label{co_dir_mu_pm}\\
D_5 \bpm \Phi^{(+)} \\ \Phi^{(-)}\epm&\equiv \left[\partial_5 -ig_5\bpm A^{(-)}_5 & A^{(+)}_5 \\
A^{(+)}_5 &A^{(-)}_5 \epm \right]\bpm \Phi^{(+)} \\ \Phi^{(-)}\epm.   \label{co_dir_5_pm}
\end{align}
It is important to note that the above action is manifestly gauge invariant under the gauge group $U(1)^\p\times U(1)$.

The next two sub-Appendices are devoted to two possible strategies for implementing spontaneous gauge symmetry breaking for this
Abelian $U(1)$ symmetric case. We are going to describe and compare: ({\it i}) SSB by vacuum expectation values of the KK modes
and ({\it ii}) SSB by a $y$-dependent vacuum expectation value of the 5D Higgs field.

\subsection{SSB by vacuum expectation values of KK modes}
\label{EWSB by vacuum expectation value of KK modes}
\noindent In this case we will choose the 5D axial gauge, $A_5^{(\pm)}=0$.
This gauge is realized by choosing the gauge parameter such that,
\beq
\Lambda^{(\pm)}(x,y)=-g_5\int dy A^{(\mp)}_5(x,y)+\hat \Lambda^{(\pm)}(x),    \label{Lambda}
\eeq
where $\hat \Lambda^{(\pm)} (x)$ is the integration constant (residual gauge freedom) and
only depends on $x^\mu$. Note that the $\hat \Lambda^{(-)}(x)$, being an odd function of $y$, must vanish.  Consequently, we are left with only one 4D gauge function, $\hat \Lambda^{(+)}(x)$.

It is convenient to parameterize the complex scalar field $\Phi^{(\pm)}(x,y)$ as
\beq
\bpm \Phi^{(+)} \\ \Phi^{(-)}\epm \equiv e^{ig_5(\pi^{(+)}\mathds{1}+\pi^{(-)}\tau_1)} \bpm \phi^{(+)} \\ \phi^{(-)}\epm, \label{H_pi_Phi}
\eeq
where $\phi^{(\pm)}(x,y)$ and $\pi^{(\pm)}(x,y)$ are real scalar fields.
We KK-decompose the scalar fields $\phi^{(\pm)}(x,y)$, $\pi^{(\pm)}(x,y)$ and the gauge fields $A^{(\pm)}_\mu(x,y)$ as
\begin{align}
\phi^{(\pm)}(x,y)&=\sum_n \phi^{(\pm)}_n(x)f^{(\pm)}_n(y),   \label{KK_Phi_pm}\\
\pi^{(\pm)}(x,y)=\sum_n \pi^{(\pm)}_n(x)a^{(\pm)}_n(y),&   \Lsp A^{(\pm)}_\mu(x,y)=\sum_n A^{(\pm)}_{\mu n}(x)a^{(\pm)}_n(y),   \label{KK_A_pm}
\end{align}
where the scalar wave-functions $f^{(\pm)}_n(y)$ and the gauge wave-functions $a^{(\pm)}_n(y)$ satisfy
\begin{align}
-\partial_5 \big(e^{4A(y)}\partial_5 f^{(\pm)}_n(y)\big)+\mu_B^2e^{4A(y)}f^{(\pm)}_n(y)&=m^{(\pm)2}_ne^{2A(y)}f^{(\pm)}_n(y),\label{eom_fn_phi_pm}\\
-\partial_5 \big(e^{2A(y)}\partial_5 a^{(\pm)}_n(y)\big)&=m_{A^{(\pm)}_n}^2a^{(\pm)}_n(y).  \label{eom_fn_A_b}
\end{align}
The wave-functions $f_n^{(\pm)}$ and $a_n^{(\pm)}$ satisfy the following orthonormality conditions,
\begin{align}
\int_{-L}^{+L}dy e^{2A(y)}f^{(\pm)}_m(y)f^{(\pm)}_n(y)=\delta_{mn}, \lsp \int_{-L}^{+L}dy a^{(\pm)}_m(y)a^{(\pm)}_n(y)=\delta_{mn}. \label{norm_condition_HA}
\end{align}
It is worth commenting here that the gauge field $A^{(\pm)}_\mu(x,y)$ and the scalar field $\pi^{(\pm)}(x,y)$ share the same $y$-dependent KK-eigen bases $a^{(\pm)}_n(y)$, as is necessitated (see below) by the  Higgs mechanism.
The KK-modes satisfy $\Box^{(4)}A^{(\pm)}_{\mu n}(x)=m_{A^{(\pm)}_n}^2A^{(\pm)}_{\mu n}(x)$.
The boundary (jump) conditions for $a^{(\pm)}_{n}(y)$ at $y=0$ and $y=\pm L$ are,
\begin{align}
\partial_5a^{(+)}_n(y)\Big\vert_{0^+}&=0,    \lsp a^{(-)}_n(y)\Big\vert_{0^+}=0,  \lsp \partial_5a^{(\pm)}_{n}(y)\Big\vert_{\pm L^\mp}=0. \label{A_bc0_b}
\end{align}
We choose the Neumann b.c. for $a^{(+)}_n(y)$ at $y=0,\pm L$ in order to insure that we get non-zero even zero-mode gauge profiles. With regard to the odd modes, we have chosen the Neumann b.c. of  $\partial_5a^{(-)}_{n}(\pm L)=0$ --- the other choice of $a_n^{(-)}(\pm L)=0$ would lead to a trivial theory with $a_n^{(-)}(y)=0$ everywhere.
Similarly, the jump and boundary conditions for the even and odd scalar profiles $f^{(+)}_n(y)$ are,
\begin{align}
\left(\partial_5 -\frac{m^2_{UV}}{2k}\right)f^{(+)}_n(y)\Big\vert_{0}=0,
\hsp \; &f^{(-)}_n(y)\Big\vert_{0}=0, \;\hsp \left(\pm\partial_5-\frac{m^2_{IR}}{k}\right)f^{(\pm)}_{n}(y)\Big\vert_{\pm L}=0.
\label{Phi_bcL_b}
\end{align}
Assuming that the KK-scale is high enough,    i.e.
$m_{KK}\sim\co(\text{few})\tev$, we can employ the effective theory where only the
lowest modes (zero-modes with masses much below $m_{KK}$) are considered.  Equation \eqref{eom_fn_A_b} along with the b.c.
\eqref{A_bc0_b} imply that the odd zero-mode wave-function of the gauge boson is zero, i.e. $a^{(-)}_{0}=0$.  As a result,
 in the effective theory the odd zero-mode
gauge boson $A^{(-)}_{0\mu}$ and odd parity Goldstone mode $\pi^{(-)}_{0}$ are not present.
In contrast, the even zero-mode wave-function for the gauge boson has a constant profile
in the bulk,    i.e.  $a^{(+)}_0=1/\sqrt{2L}$. The forms of the scalar zero-mode wave functions $f^{(\pm)}_0(y)$ are given by:
\beq
f^{(\pm)}_0(|y|)\simeq\sqrt{k(1+\beta)}e^{kL} e^{(2+\beta)k(|y|-L)},  \label{f0pm}
\eeq
where, $f^{(-)}_0(y)=\epsilon(y)f^{(-)}_0(|y|)$.
The low-energy effective action for the zero-modes is
\begin{align}
S_{Ab}^{eff}=-\int d^4x &\bigg\{\frac14 F^{0(+)}_{\mu\nu}F_{0(+)}^{\mu\nu}+ \partial_\mu \phi^{(+)}_0\partial^\mu \phi^{(+)}_0 -\mu^{2}\phi^{(+)2}_0 + \partial_\mu \phi^{(-)}_0\partial^\mu \phi^{(-)}_0 -\mu^{2}\phi^{(-)2}_0  \notag\\
&+\Big(g^{(+)2}_{0000}\phi^{(+)2}_0+\bar g^{(+)2}_{0000}\phi^{(-)2}_0\Big)\Big(A^{(+)}_{0\mu}-\partial_\mu\pi^{(+)}_{0}\Big)^2\notag\\
&+\lambda^{(+)}_{0000}\phi^{(+)4}_0+\lambda^{(-)}_{0000}\phi^{(-)4}_0 +6\lambda_{0000}\phi^{(+)2}_0\phi^{(-)2}_0 \bigg\},    \label{eff_action_Abelian_zm}
\end{align}
where the mass parameter $\mu$ is defined as $\mu^{2}\equiv(1+\beta)m_{KK}^2\delta_{IR}$, with the parameters
\beq
\delta_{IR}\equiv\frac{m^2_{IR}}{k^2}-2(2+\beta),  \lsp m_{KK}\equiv ke^{-kL} \hsp\text{and}\hsp\beta\equiv\sqrt{4+\mu^2_B/k^2}.
\label{delta_IR_mKK}
\eeq
Above, the couplings have the following form in terms of the parameters of the 5D theory:
\begin{align}
\lambda^{(\pm)}_{0000}&=\lambda_{0000}\simeq\lambda\equiv\lambda_{IR}(1+\beta)^2,\Lsp
g^{(+)}_{0000}= \bar g^{(+)}_{0000}= g_4\equiv\frac{g_5}{\sqrt{2L}}.
\end{align}
Our effective theory can be parameterized by redefining $A_{0\mu}^{(+)}(x)\equiv A_\mu(x)$, $\pi^{(+)}_0(x)\equiv \pi(x)$,
\beq
\Phi_1(x)\equiv e^{ig_4 \pi(x)}\phi^{(+)}_0(x), \lsp \Phi_2(x)\equiv e^{ig_4 \pi(x)}\phi^{(-)}_0(x),  \label{H_12_ab}
\eeq
in which case the above effective action can be written in a nice gauge covariant form as
\begin{align}
S_{Ab}^{eff}=-\int d^4x \bigg\{\frac14 F_{\mu\nu}F^{\mu\nu}+ {\cal D}_\mu \Phi^{\ast}_1{\cal D}^\mu \Phi_1+ {\cal D}_\mu \Phi^{\ast}_2{\cal D}^\mu \Phi_2 +V(\Phi_1,\Phi_2) \bigg\},    \label{eff_action_ab_gauge_cov}
\end{align}
where the covariant derivative is defined as ${\cal D}_\mu \equiv \partial_\mu-ig_4 A_\mu$ and the scalar potential can be written as
\begin{align}
V(\Phi_1,\Phi_2) =&-\mu^{2}|\Phi_1|^2 -\mu^{2}|\Phi_2|^2 +\lambda|\Phi_1|^4 +\lambda|\Phi_2|^4\notag+6\lambda|\Phi_1|^2|\Phi_2|^2.    \label{potenial_ab}
\end{align}

It is important to note that, after choosing the gauge $A_5(x,y)=0$, we are left with a residual
gauge freedom with a single purely 4D gauge parameter $\hat \Lambda^{(+)}(x)$ such that the above Lagrangian is
invariant under the $U(1)$ gauge transformation,
\begin{align}
A_\mu(x)\to A_\mu^{\prime}(x)&=A_\mu(x)+\frac{1}{g_4}\partial_\mu \hat \Lambda^{(+)}(x),   \\
\Phi_1(x)\to \Phi^{\prime}_1(x)=e^{i g_4 \hat \Lambda^{(+)}}\Phi_1(x),&    \lsp \Phi_2(x)\to \Phi^{\prime}_2(x)=e^{i g_4 \hat \Lambda^{(+)}}\Phi_2(x).
\end{align}
\begin{wrapfigure}{r}{0.4\textwidth}
\vspace{-23pt}
\centering\includegraphics[scale=0.57]{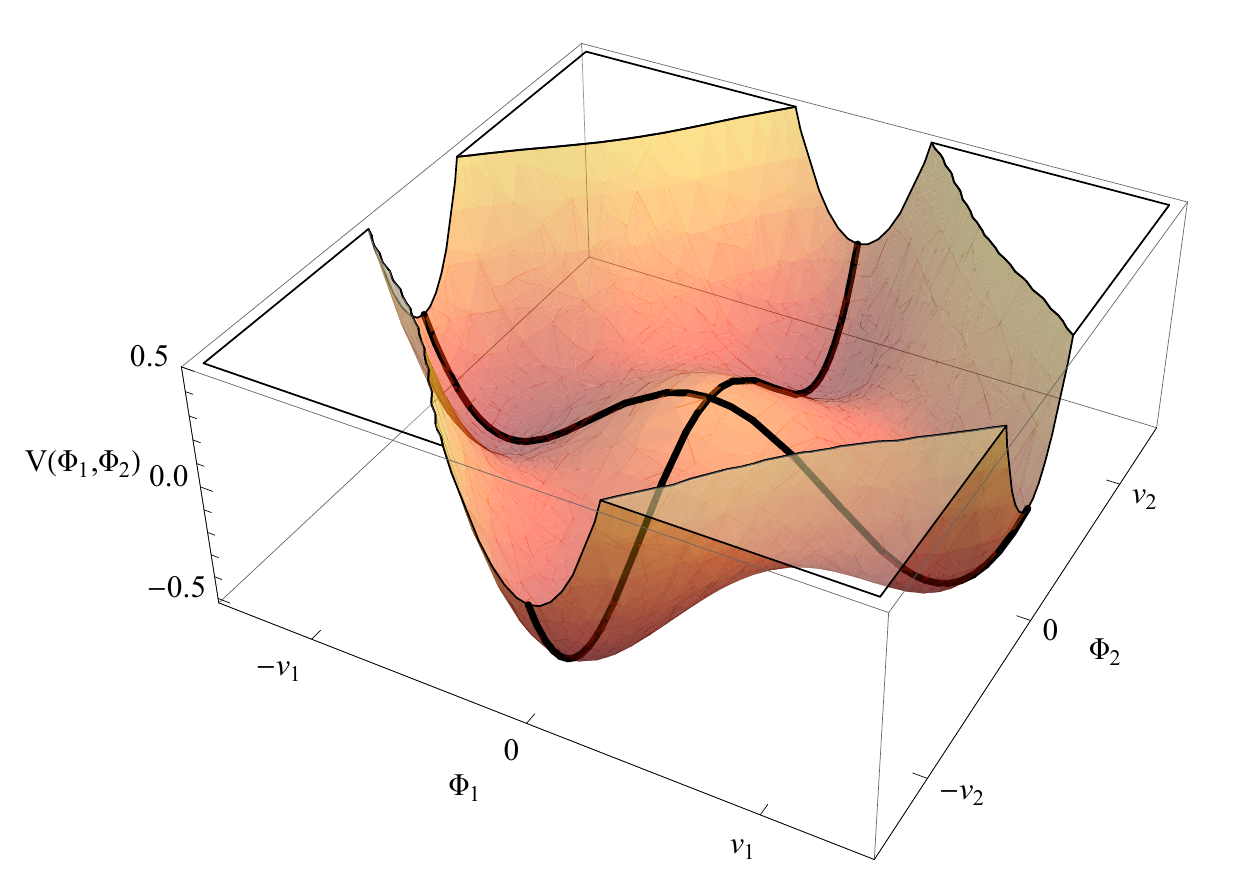}
\caption{This graph illustrates shape of the scalar potential $V(\Phi_1,\Phi_2)$ as a
function of the fields $\Phi_1$ and $\Phi_2$.}
\label{fig_V}
\end{wrapfigure}
Thus, besides $\mathbb{Z}_2^\prime \times \mathbb{Z}_2$ symmetry, the above potential is invariant under $U(1)^\prime \times U(1)$. One $U(1)$
has been gauged while the other is a remnant of the global unbroken symmetry associated with the odd gauge transformation ($\Lambda^{(-)}$)
defined in Eqs.~\eqref{gauge_trans_A_5}-\eqref{H_pm_transformation}.

One can easily see from Fig. \ref{fig_V} that the scalar potential $V(\Phi_1,\Phi_2)$
has four degenerate global minima at $(\pm v_1,0)$ and $(0,\pm v_2)$. One can choose
any of these global vacua. We select the vacuum such that the even parity Higgs $\Phi_1$ acquires a vev, whereas the odd parity Higgs $\Phi_2$
does not. That choice of vacuum implies values of $v_1$ and $v_2$ given by,
\beq
v_1=\frac{\mu}{\sqrt\lambda},  \Lsp v_2=0.        \label{v1_v2_ab}
\eeq
Now let us consider the fluctuations of the above fields around our choice of the vacuum,
\beq
\Phi_1(x)=\frac{1}{\sqrt2}\Big(v_1+h\Big)e^{ig_4 \pi(x)},    \Lsp \Phi_2(x)=\frac{1}{\sqrt2} \chi e^{ig_4 \pi(x)} .    \label{H1_H2_def}
\eeq
We rewrite our effective action \eqref{eff_action_ab_gauge_cov} only up to the quadratic order in fluctuations as
\begin{align}
S_{Ab}^{(2)}=&-\int d^4x \bigg\{\frac14 F_{\mu\nu}F^{\mu\nu}+ \frac{g^{2}_{4}v_1^2}{2} \Big(A_\mu  -\partial_\mu\pi\Big)^2+\frac12\partial_\mu h\partial^\mu h+ \frac12m^2_h h^2+\frac12\partial_\mu \chi\partial^\mu \chi+ \frac12m^2_\chi \chi^2\bigg\}.    \label{eff_action_ab_quadratic_0}
\end{align}
The mixing between $A_\mu$ and $\pi$ in the above action can be removed by an appropriate 4D gauge choice. Here we will choose the 4D unitary gauge such that $\pi=0$ and the gauge field acquires mass. The remaining scalars are $h$ and $\chi$ with masses
\beq
m^2_h=m^2_\chi=2\mu^2.         \label{masses_ab}
\eeq
Hence, the full effective Abelian action can be written in the 4D unitary gauge as
\begin{align}
S_{Ab}^{eff}=-\int d^4x &\bigg\{\frac14 F_{\mu\nu}F^{\mu\nu}+ \frac12 m_A^2A_\mu A^\mu +\frac12\partial_\mu h\partial^\mu h+ \frac12m^2_h h^2+\frac12\partial_\mu \chi\partial^\mu \chi+ \frac12m^2_\chi \chi^2   \notag\\
&+\lambda v_1 h\Big(h^2+3\chi^2\Big)+ \frac14 \lambda h^4+ \frac14 \lambda\chi^4+\frac32\lambda h^2\chi^2   \notag\\
&+g_4^2 v_1hA_\mu A^\mu+\frac12g_4^2\Big(h^2+\chi^2\Big)A_\mu A^\mu\bigg\},    \label{eff_action_ab_full}
\end{align}
where
\beq
m_A^2\equiv g_4^2v_1^2=g_4^2\frac{\mu^2}{\lambda}.          \label{mass_A}
\eeq
To summarize, the zero-mode effective theory for the Abelian case has two real scalars with equal mass and a massive gauge boson. Also, the above action is invariant under the $\mathbb{Z}_2$ symmetry $h\to h$ and $\chi\to-\chi$.

\subsection{SSB by a vacuum expectation value of the 5D Higgs field}
\label{EWSB by vacuum expectation value of 5D Higgs field}
\noindent In this subsection, we write the complex scalar fields $\Phi^{(\pm)}$ as
\begin{align}
\bpm \Phi^{(+)}(x,y)\\\Phi^{(-)}(x,y) \epm&=\frac{1}{\sqrt2}e^{ig_5(\pi^{(+)}(x,y)\mathds{1}+\pi^{(-)}(x,y)\tau_1)}\bpm v(y)+h^{(+)}(x,y)\\ h^{(-)}(x,y)\epm. \label{higgs_ev}
\end{align}
As mentioned above, the vev $v(y)$ is only associated with the even Higgs field $\Phi^{(+)}$ and it is even under the geometric $\mathbb{Z}_2$ parity~\footnote{There exists also an odd parity vacuum solution but as our geometric setup is symmetric, therefore we choose the even vacuum solution $v(y)$ for the scalar field. With this choice of the background solution, the fluctuations will have a definite parity. Note that the choice of odd vacuum solution could lead to the breaking of geometric $\mathbb{Z}_2$ symmetry.}. The fluctuations $h^{(+)}(x,y)$ and $\pi^{(+)}(x,y)$ are even, whereas the fluctuations $h^{(-)}(x,y)$ and $\pi^{(-)}(x,y)$ are odd under the $\mathbb{Z}_2$ geometric parity.

We can write the action Eq.~\eqref{action_5d_pm} up to quadratic order in fields as
\begin{align}
S_{Ab}^{(2)}=&-\int d^5x\bigg\{\frac14 F^{(+)}_{\mu\nu}F^{(+)\mu\nu}+\frac12 e^{2A(y)}\Big((\partial_\mu A^{(+)}_5)^2+(\partial_5 A^{(+)}_\mu)^2+g_5^2v^{2}A^{(+)}_\mu A^{(+)\mu}\Big)  \notag\\
&+\Big(e^{2A(y)}g_5^2v^2\pi^{(+)}-\partial_5(e^{2A(y)}A^{(-)}_5)\Big)\partial_\mu A^{(+)\mu} +\frac12e^{2A(y)}\Big( (\partial_\mu h^{(+)})^2+g_5^2v^2(\partial_\mu \pi^{(+)})^2\Big)\notag\\
&+\frac12e^{4A(y)}\Big((\partial_5 v+\partial_5 h^{(+)})^2+g^2_5v^{2}\Big(A^{(-)}_5-\partial_5 \pi^{(+)}\Big)^2+\mu_B^2 (v+h^{(+)})^2\Big)   \notag\\
&+\frac14 F^{(-)}_{\mu\nu}F^{(-)\mu\nu}+\frac12 e^{2A(y)}\Big((\partial_\mu A^{(-)}_5)^2+(\partial_5 A^{(-)}_\mu)^2+g^2_5v^{2}A^{(-)}_\mu A^{(-)\mu}\Big)  \notag\\
&+\Big(e^{2A(y)}g_5^2v^2\pi^{(-)}-\partial_5(e^{2A(y)}A^{(+)}_5)\Big)\partial_\mu A^{(-)\mu} +\frac12e^{2A(y)}\Big(( \partial_\mu h^{(-)})^2+g_5^2v^2(\partial_\mu \pi^{(-)})^2\Big)\notag\\
&+\frac12e^{4A(y)}\Big((\partial_5h^{(-)})^2+g^2_5 v^2\Big(A^{(+)}_5-\partial_5 \pi^{(-)}\Big)^2+\mu_B^2 h^{(-)2}\Big)\bigg\}.   \label{action_2_pm}
\end{align}
where the indices are raised and lowered by the Minkowski metric. The bulk equation of motion for the background Higgs vev corresponding to the above action is
\begin{align}
&\bigg[-\partial_5\left(e^{4A(y)}\partial_5\right)+\mu_B^2 e^{4A(y)}\bigg]v(y)   \notag\\
&=-e^{4A(y)}\left[\frac{\partial V_{IR}(v)}{\partial v}\delta(y+L)+\frac{\partial V_{UV}(v)}{\partial v}\delta(y)+\frac{\partial V_{IR}(v)}{\partial v}\delta(y-L)\right],    \label{eom_higgs_vev}
\end{align}
and the bulk equations of motion for all the fluctuations are
\begin{align}
&\bigg[-e^{2A(y)}\Box^{(4)} -\partial_5\left(e^{4A(y)}\partial_5\right)+\mu_B^2 e^{4A(y)}\bigg]h^{(\pm)}(x,y)  \notag\\
&=-e^{4A(y)}\left[\frac{\partial^2 V_{UV}(v)}{\partial v^2}h^{(\pm)}\delta(y)+\frac{\partial^2 V_{IR}(v)}{\partial v^2}h^{(\pm)}\Big(\delta(y-L)+\delta(y+L)\Big)\right],\label{eom_h_ab}\\
&\Box^{(4)}A^{(\pm)}_\mu+\partial_5\Big(M_A^2\partial_5A^{(\pm)}_\mu\Big)-M_A^2A^{(\pm)}_\mu \notag\\
&=\partial_\mu\Big(\partial^\nu A^{(\pm)}_\nu+\partial_5(e^{2A(y)}A^{(\mp)}_5)-M_A^2\pi^{(\pm)}\Big),         \label{eom_A_mu_p}\\
&\Box^{(4)}A^{(\pm)}_5-\partial_5\Big(\partial^\nu A^{(\mp)}_\nu\Big)-M_A^2\Big(A^{(\pm)}_5-\partial_5\pi^{(\mp)}\Big)=0,         \label{eom_A_5_p}\\
&\Box^{(4)}\pi^{(\pm)}-\partial^\nu A^{(\pm)}_\nu-M_A^{-2}\partial_5\Big(M_A^2e^{2A(y)}(A^{(\mp)}_5-\partial_5\pi^{(\pm)})\Big)=0,  \label{eom_sigma}
\end{align}
where $M_A^2\equiv g^2_5v^{2}(y)e^{2A(y)}$. The jump  conditions at the UV-brane following from the equations of motion above are:
\beq
\left(\partial_5-\frac{\partial V_{UV}(v)}{\partial v}\right) v(y)\Big\vert_{0^+}=0, \lsp\left(\partial_5-\frac{\partial^2V_{UV}(v)}{\partial v^2}\right) h^{(+)}(x,y)\Big\vert_{0^+}=0,
\label{boundary_cond_v_ab}
\eeq
whereas the odd fields must vanish at $y=0$.
In addition, we choose the boundary conditions at $\pm L$:
\begin{align}
 \left(\pm\partial_5+\frac{\partial V_{IR}(v)}{\partial v}\right) v(y)\Big\vert_{\pm L^\mp}&=0,
 \hsp \left(\pm\partial_5+\frac{\partial^2V_{IR}(v)}{\partial v^2}\right) h^{(\pm)}(x,y)\Big\vert_{\pm L^\mp}=0,  \label{boundary_cond_h_ab}\\
\partial_\mu A^{(\pm)}_5(x,y)-\partial_5A^{(\mp)}_\mu(x,y)\Big\vert_{\pm L^\mp}&=0,  \lsp A^{(\mp)}_5(x,y)-\partial_5\pi^{(\pm)}(x,y)\Big\vert_{\pm L^\mp}=0,    \label{boundary_cond_gauge}
\end{align}
where $L^\pm\equiv L\pm\epsilon$ for $\epsilon\to0$.

We find the following vacuum solutions for $v(y)$:
\begin{align}
v(y)&=C_1 e^{(2+\beta)k|y|}+C_2e^{(2-\beta)k|y|},  &&\lsp -L\leq y\leq L,    \label{v_sol_tp}
\end{align}
where $C_1$ and $C_2$ are the integration constants. We apply the jump condition \eqref{boundary_cond_v_ab} at $y=0$ and the boundary condition at $y=\pm L$ \eqref{boundary_cond_h_ab} to fix the two integration constants as,
\begin{align}
C_2&=-\frac{\delta_{UV}}{\delta_{UV}+4\beta}C_1,  \label{B_constant_t}\\
C_1&=\sqrt{\frac{k^3}{\lambda_{IR}}\left(\delta_{IR}-\frac{\delta_{UV}(\delta_{IR}+4\beta)}{\delta_{UV}+4\beta}e^{-2\beta kL}\right)}e^{-(2+\beta)kL} \left(1-\frac{e^{-2\beta kL}\delta_{UV}}{\delta_{UV}+4\beta}\right)^{-3/2},
\label{Ap_constant_t}
\end{align}
where $\delta_{UV}\equiv m_{UV}^2/k^2-2(2+\beta)$ and $\delta_{IR}$ is defined in Eq. \eqref{delta_IR_mKK}.
For $kL\gg1$ and $\beta>0$, the terms proportional to $e^{-2\beta kL}$ are negligible in the region of interest
(near the IR-brane). Hence the vacuum solution for the scalar field can be written as:
\beq
v(y)\simeq\sqrt{\frac{k^3\delta_{IR}}{\lambda_{IR}}}e^{(2+\beta)k(|y|-L)} \equiv v_4 f_v(y),    \label{vp_sol_toy}
\eeq
where the constant vev $v_4$ and the $y$-dependent vev profile $f_v(y)$ are:
\beq
v_4\equiv\sqrt{\frac{m_{KK}^2\delta_{IR}}{\lambda_{IR}(1+\beta)}}, \lsp f_v(y)\equiv \sqrt{k(1+\beta)}e^{kL}e^{(2+\beta)k(|y|-L)}.    \label{v4_fv_sol}
\eeq
The $y$-dependent vev profile satisfies the orthonormality condition
\beq
\int_{-L}^{L}dy e^{2A(y)}f_v^2(y)=1.        \label{f_v_norm}
\eeq
From the above solution we conclude that for $\lambda_{IR}>0$ (as required by the positivity of the tree-level potential)
one needs $\delta_{IR}>0$, i.e. $m_{IR}^2/k^2>2(2+\beta)$.
It is worth mentioning here that the quartic term in the IR-brane potential is crucial for a
non-trivial vev profile $v^{(\pm)}(y)$. If the quartic term would have been absent in the $V_{IR}$, i.e.  $\lambda_{IR}=0$, then the b.c. \eqref{boundary_cond_v_ab} and \eqref{boundary_cond_h_ab} would have implied $v(y)=0$. Even though the quartic term is not in the bulk (only localized at the IR-brane), nevertheless, the b.c. imply the non-zero profile in the bulk.

In the above action there are mixing terms of $\partial_\mu A^{(\pm)\mu}$ with the scalars
$\pi^{(\pm)}$ and $A_5^{(\pm)}$,
which can be canceled by adding the following gauge fixing Lagrangian to the above action,
\begin{align}
S_{GF}=-\int d^5x&\bigg\{\frac{1}{2\xi} \left[\partial_\mu A^{\mu(+)}-\xi\left(M_A^2\pi^{(+)}-\partial_5\Big(e^{2A(y)}A^{(-)}_5\Big)\right)\right]^2    \notag \\
&+\frac{1}{2\xi} \left[\partial_\mu A^{\mu(-)}-\xi\left(M_A^2\pi^{(-)}-\partial_5\Big(e^{2A(y)}A^{(+)}_5\Big)\right)\right]^2\bigg\}.  \label{S_GF}
\end{align}
One can identify the Goldstone modes from the above two Eqs.~\eqref{action_2_pm} and \eqref{S_GF}:
\beq
\Pi^{(\pm)}(x,y)\equiv M_A^2\pi^{(\pm)}-\partial_5\big(e^{2A(y)}A^{(\mp)}_5\big),  \label{goldstone_G_pm}
\eeq
along with the two pseudoscalars ${\cal A}_5^{(\pm)}(x,y)$ given as
\beq
{\cal A}_5^{(\pm)}(x,y)\equiv A^{(\pm)}_5-\partial_5 \pi^{(\mp)}.  \label{Phi_pm}
\eeq
The resulting four pseudoscalars above along with the two $h^{(\pm)}$ scalar fields agrees with the naive counting before SSB of three even-parity
scalars ($h^{(+)}(x,y)$, $\pi^{(+)}(x,y)$ and $A^{(+)}_5(x,y)$) and three odd-parity scalars
($h^{(-)}(x,y)$, $\pi^{(-)}(x,y)$ and $A^{(-)}_5(x,y)$). It is seen from the Eq.~(\ref{action_2_pm}) quadratic action
that both the even and odd gauge bosons $A^{(\pm)}_\mu(x,y)$ acquire mass from the Higgs mechanism,
whereby the  two Goldstone bosons are eaten up by these gauge bosons.

In order to obtain an effective 4D Lagrangian we need to integrate the above quadratic Lagrangian over the $y$-coordinate.
The first step to achieve this is to decompose all the fields in KK-modes. We will use the following decomposition,
\begin{align}
A^{(\pm)}_\mu(x,y)&=\sum_n A^{(\pm)}_{\mu n}(x) \tilde a^{(\pm)}_n(y),      \lsp &&h^{(\pm)}(x,y)=\sum_n h^{(\pm)}_n\tilde f^{(\pm)}_n(y), \label{KK-mode_AH} \\
\Pi^{(\pm)}(x,y)&=\sum_n \Pi^{(\pm)}_n(x) \tilde a^{(\pm)}_n(y) \tilde m^{(\pm)}_{A_n}, \lsp &&{\cal A}_5^{(\pm)}(x,y)=\sum_n {\cal A}^{(\pm)}_n(x) \eta^{(\pm)}_n(y), \label{KK-mode_phi_sigma}
\end{align}
where $\tilde a^{(\pm)}_n(y)$, $\eta^{(\pm)}_n(y)$ and $\tilde f^{(\pm)}_n(y)$ are the 5D profiles for the vector fields (the same for the Goldstone fields), the pseudoscalars and the Higgs bosons, respectively. The e.o.m. for the wave-functions $\tilde f^{(\pm)}_n(y)$, $\tilde a^{(\pm)}_n(y)$ and $\eta^{(\pm)}_n(y)$ are
\begin{align}
-\partial_5 (e^{4A(y)}\partial_5 \tilde f^{(\pm)}_n(y))+\mu_B^2e^{4A(y)}\tilde f^{(\pm)}_n(y)&= \tilde m^{(\pm)2}_ne^{2A(y)}\tilde f^{(\pm)}_n(y),  \label{eom_fn_phi}\\
-\partial_5 (e^{2A(y)}\partial_5\tilde  a^{(\pm)}_n(y))+M_A^2\tilde a^{(\pm)}_n(y)&=\tilde m_{A_n}^{(\pm)2}\tilde a^{(\pm)}_n(y),  \label{a_n_eom}\\
-\partial_5 \left(M_A^{-2}\partial_5(M_A^2e^{2A(y)}\eta^{(\pm)}_n(y))\right)+M_A^2\eta^{(\pm)}_n(y)&=m_{{\cal A}_n}^{(\pm)2}\eta^{(\pm)}_n(y),  \label{eta_n_eom}
\end{align}
where $\tilde m_n^{(\pm)}$, $\tilde m_{A_n}^{(\pm)}$ and $m_{{\cal A}_n}^{(\pm)}$ are the KK-masses for $h^{(\pm)}_n$, $A^{(\pm)}_{\mu n}(x)$ and $\phi^{(\pm)}_n(x)$. The normalization conditions for the wave-functions $\tilde f^{(\pm)}_n(y)$, $\tilde a_n(y)$ and $\eta_n(y)$ are
\begin{align}
\int_{-L}^{+L}dy e^{2A(y)}\tilde f^{(\pm)}_m(y)\tilde f^{(\pm)}_n(y)&=\delta_{mn},  \lsp \int_{-L}^{+L}dy \tilde a^{(\pm)}_m(y)\tilde a^{(\pm)}_n(y)=\delta_{mn}    \label{norm_fn}\\
\int_{-L}^{+L}dy &\frac{M_A^2e^{2A(y)}}{m_{{\cal A}_m}^{(\pm)}m_{{\cal A}_n}^{(\pm)}}\eta^{(\pm)}_m(y)\eta^{(\pm)}_n(y)=\delta_{mn}. \label{norm_condition_eta_n}
\end{align}
Following the general strategy mentioned in Sec. \ref{KK-Parity}, we choose the $y=0$ b.c. for the even wave functions as Neumann (or mixed) b.c., whereas
{\it all} the odd-mode wave functions satisfy Dirichlet b.c. at $y=0$:
\begin{align}
\left(\partial_5 -\frac{m^2_{UV}}{2k}\right)\tilde f^{(+)}_n(y)\Big\vert_{0}&=0,    \lsp \; \; \tilde f^{(-)}_n(y)\Big\vert_{0}=0,  \label{Phi_bc0}\\
\partial_5 \tilde a^{(+)}_n(y)\Big\vert_{0}=0, \lsp \tilde a^{(-)}_n(y)\Big\vert_{0}&=0,  \lsp \partial_5\eta^{(+)}_n(y)\Big\vert_{0}=0, \lsp \eta^{(-)}_n(y)\Big\vert_{0}=0. \label{bc_a_0}
\end{align}
The b.c. for wave-functions $\tilde f^{(\pm)}_n$, $\tilde a^{(\pm)}_{n}$ and $\eta^{(\pm)}_n$ at $y=\pm L$ follow from Eqs. \eqref{boundary_cond_h_ab}-\eqref{boundary_cond_gauge},
\begin{align}
\left(\pm\partial_5-\frac{m^2_{IR}}{2k}+\frac{3\lambda_{IR}}{2k^2}v^2(y)\right)\tilde f^{(\pm)}_{n}(y)\Big\vert_{\pm L^\mp}=0, \hsp
\partial_5 \tilde a^{(\pm)}_n(y)\Big\vert_{\pm L}&=0,   \hsp \eta^{(\pm)}_n(y)\Big\vert_{\pm L}=0.  \label{bc_a_L_eta}
\end{align}
One can also easily find the KK-decomposition of the fluctuation fields $A^{(\pm)}_5(x,y)$ and $\pi^{(\pm)}(x,y)$ in terms of Goldstone bosons $\Pi^{(\pm)}$ and the physical scalars ${\cal A}_5^{(\pm)}$ from
Eqs.~\eqref{goldstone_G_pm}-\eqref{Phi_pm}:
\begin{align}
A^{(\pm)}_5(x,y)&=\sum_n \left(\frac{\Pi^{(\mp)}_n(x) }{\tilde m^{(\mp)}_{A_n}}\partial_5\tilde a^{(\mp)}_n(y)-\frac{M_A^2}{(m^{(\pm)}_{{\cal A}_n})^2}{\cal A}^{(\pm)}_n(x) \eta^{(\pm)}_n(y)\right),  \label{KK-mode_A_5}\\
\pi^{(\pm)}(x,y)&=\sum_n \left(\frac{\Pi^{(\pm)}_n(x)}{\tilde m^{(\pm)}_{A_n}} \tilde a^{(\pm)}_n(y)-\frac{M_A^{-2}}{(m^{(\mp)}_{{\cal A}_n})^2}\partial_5\left(M_A^2e^{2A(y)}\eta^{(\mp)}_n(y)\right){\cal A}^{(\mp)}_n(x)\right),  \label{KK-mode_pi}
\end{align}

Now we consider the low-energy effective theory obtained by assuming that the KK-mass scale is high enough  that we can integrate out
all the heavier KK modes and keep only the zero-modes of the theory. From here on, we choose the unitary
gauge such that $\xi\to\infty$ which implies $\Pi^{(\pm)}_n(x)\to 0$. Moreover, with our choice of boundary
conditions for $a^{(-)}_{0}(y)$ and $\eta^{(\pm)}_0(y)$ in Eqs. \eqref{bc_a_0} and \eqref{bc_a_L_eta} one can
see that the corresponding wave-functions for zero-modes are vanishing, i.e.  there will be no zero-modes
$A^{(-)}_{0\mu}(x)$ and ${\cal A}^{(\pm)}_0(x)$ in our effective theory. The $y$-dependent vev
and the zero-mode profiles for even and odd Higgs are given by
\begin{align}
f_v(y)&\equiv\sqrt{k(1+\beta)}e^{kL}e^{(2+\beta)k(|y|-L)},   \label{vp_sol_ab}\\
\tilde f^{(\pm)}_0(|y|)\approx\sqrt{k(1+\beta)}&e^{kL} e^{(2+\beta)k(|y|-L)}, \lsp \tilde f^{(-)}_0(y)=\epsilon(y)\tilde f^{(-)}_0(|y|), \label{f0pm_ab}
\end{align}
where $\mu^{2}\equiv(1+\beta)m_{KK}^2\delta_{IR}$ and $\lambda\equiv\lambda_{IR}(1+\beta)^2$.
It is important to comment here that at the leading order the vev profile  and zero-mode profiles are the same.  However, there are finite corrections which are suppressed by $\co \left(m_h^2/m_{KK}^2\right)$ as given below
\beq
\frac{\tilde f^{(\pm)}_0(|y|)}{f_v(|y|)}=1+\frac{m_h^2}{m_{KK}^2}\left(\frac{1-e^{2k(|y|-L)}}{4(1+\beta)}+\co\Big(\frac{m_h^2}{m_{KK}^2}\Big)\right).  \label{f0_fv_ratio}
\eeq
We can now write down the effective theory for the zero-modes in the unitary gauge:
\begin{align}
S_{eff}=-\int d^4x &\bigg\{\frac14 F_{\mu\nu}F^{\mu\nu} +\frac12 \tilde m^2_A A_\mu A^\mu+\frac12\partial_\mu h\partial^\mu h+ \frac12\tilde m^2_h h^2  +\frac12\partial_\mu \chi\partial^\mu \chi+ \frac12\tilde m^2_\chi \chi^2 \notag\\
&+ \frac14 \lambda h^4+ \frac14 \lambda\chi^4+\frac32\lambda h^2\chi^2 +\lambda v_4h\Big( h^2+3\chi^2\Big)\notag\\
&+\tilde g_4^2v_4hA_\mu A^\mu  +\frac12\tilde g_4^2\Big(h^2+\chi^2\Big)A_\mu A^\mu\bigg\},   \label{S_eff_ab}
\end{align}
where we have denoted $A^{(+)}_{0\mu}(x)\equiv A_\mu(x)$ and we have suppressed the
zero-mode subscript `$0$' for all modes. After some algebra, using the boundary conditions, one can  find the masses of the zero-mode scalars and gauge boson at the leading order:
\beq
\tilde m_{h}^2=\tilde m_{\chi}^2\simeq2\mu^2,     \Lsp\tilde m_{A}^2\simeq\frac{1}{2L}\int_{-L}^{L}dyM_A^2=\tilde g^2_4v^2_4,      \label{scalar_mass_ab}
\eeq
where $\mu^2\equiv (1+\beta)m_{KK}^2\delta_{IR}$, $v_4\equiv \mu/\lambda$ and  $\tilde g_4\equiv g_5/\sqrt{2L}$.

\paragraph{Comparison:} In order to facilitate comparison between the two approaches,
 we collect information concerning all the low-energy degrees of freedom for
both pictures in Table~\ref{comparison}. Comparing the effective theories obtained within EWSB induced by the Higgs KK-mode vev and by a 5D-Higgs vev in \eqref{eff_action_ab_full} and \eqref{S_eff_ab} one finds that  both approaches give exactly the {\it same} zero-mode effective theory up to $\co(m_h^2/m_{KK}^2\sim10^{-3})$ corrections. We have checked that the scalar masses are exactly same to all orders in the expansion parameter $m_h^2/m_{KK}^2$. In contrast, the gauge boson masses and the couplings can have subleading differences of order $\co(m_h^2/m_{KK}^2)$. Note that we have neglected all the effects due to the non-zero KK-modes, such effects being suppressed by their masses,    i.e.  $\co(m_h^2/m_{n}^2)$. Hence we conclude that the two approaches  to EWSB discussed above give the same low-energy (zero-mode) effective theory aside from small deviations of order $\co(m_h^2/m_{KK}^2)$. To make this comparison more transparent we summarize the parameters of both effective theories in terms of the fundamental parameters of the 5D theory in Table \ref{comp_para}.
The observed agreement at the zero-mode level of the effective theory is a non-trivial verification of the results obtained here.
\begin{table}
\centering
{\tabulinesep=1pt
\rowcolors{1}{agray}{white}
\begin{tabu}{|[1pt]l|[0.75pt]c|[0.75pt]c|[0.75pt]c|[0.75pt]c|[0.75pt]c|[0.75pt]c|[1pt]}\tabucline[1pt]{-}
&\multicolumn{3}{c|[0.75pt]}{EWSB by KK mode vev} & \multicolumn{3}{c|[1pt]}{EWSB by 5D Higgs vev} \\
\tabucline[1pt]{-}
5D fields  &KK-modes& $n=0$& $n\neq0$&KK-modes& $n=0$& $n\neq0$\\
\tabucline[1pt]{-}
$\text{Re}{\Phi^{(+)}}$ &$\phi^{(+)}_n(x)$& \cmark& \cmark&$h^{(+)}_n(x)$&\cmark& \cmark\\
$\text{Re}{\Phi^{(-)}}$ &$\phi^{(-)}_n(x)$& \cmark& \cmark&$h^{(-)}_n(x)$&\cmark& \cmark\\
$\text{Im}{\Phi^{(+)}}$ &$\pi^{(+)}_n(x)$& \xmark (4D g.c.)& \cmark&$\Pi^{(+)}_n(x)$&\xmark (4D g.c.)& \xmark (4D g.c.)\\
$\text{Im}{\Phi^{(-)}}$ &$\pi^{(-)}_n(x)$& \xmark (b.c.)& \cmark&$\Pi^{(-)}_n(x)$&\xmark (4D g.c.)& \xmark (4D g.c.)\\
$A^{(+)}_5$ &$A^{(+)}_{5n}(x)$& \xmark (5D g.c.)& \xmark (5D g.c.)&${\cal A}^{(+)}_n(x)$&\xmark (b.c.)& \cmark\\
$A^{(-)}_5$ &$A^{(-)}_{5n}(x)$& \xmark (5D g.c.)& \xmark (5D g.c.)&${\cal A}^{(-)}_n(x)$&\xmark (b.c.)& \cmark\\
$A^{(+)}_\mu$ &$A^{(+)}_{\mu n}(x)$& \cmark& \cmark&$A^{(+)}_{\mu n}(x)$&\cmark& \cmark\\
$A^{(-)}_\mu$ &$A^{(-)}_{\mu n}(x)$& \xmark (b.c.)& \cmark&$A^{(-)}_{\mu n}(x)$&\xmark (b.c.)& \cmark\\
\tabucline[1pt]{-}
\end{tabu}}
\caption{Comparison of dynamical d.o.f. between KK-mode-vev and 5D-Higgs-vev EWSB. The b.c. (boundary condition)
and g.c. (gauge choice) show why a given mode is not present in the corresponding effective theory. Note that $\Pi^{(\pm)}_n$ is a mixture of
$\pi^{(\pm)}_n$ and $A_{5n}^{(\pm)}$, see Eq. \eqref{goldstone_G_pm}.}
\label{comparison}
\end{table}
\begin{table}
\centering
{\tabulinesep=1pt
\rowcolors{1}{agray}{white}
\begin{tabu}{|[1pt]l|[0.75pt]l|[0.75pt]c|[1pt]}\tabucline[1pt]{-}
EWSB by KK mode vev& EWSB by 5D Higgs vev& Comment\\
\tabucline[1pt]{-}
$f_0(y)\simeq\sqrt{k(1+\beta)}e^{kL}e^{(2+\beta)k(|y|-L)}$ &$\tilde f_0(y)\simeq\sqrt{k(1+\beta)}e^{kL}e^{(2+\beta)k(|y|-L)}$& same\\
$v_1^{2}=\frac{\mu^2}{\lambda}\left(1-\co\Big(\frac{m_h^2}{m_{KK}^2}\Big)\right)$&$v_4^{2}=\frac{\mu^2}{\lambda}$ \hsp and \hsp $f_v(y)=\tilde f_0(y)$ & $\co\Big(\frac{m_h^2}{m_{KK}^2}\Big)$\\
$m_h^{2}=m_\chi^2=2\mu^2\left(1-\co\Big(\frac{m_h^2}{m_{KK}^2}\Big)\right)$&$\tilde m_h^{2}=\tilde m_\chi^2=2\mu^2\left(1-\co\Big(\frac{m_h^2}{m_{KK}^2}\Big)\right)$ & same\\
$a_0(y)=\frac{1}{\sqrt{2L}}$&$\tilde a_0(y)=\frac{1}{\sqrt{2L}}\left(1+\co\Big(\frac{m_h^2}{m_{KK}^2}\Big)\right)$ & $\co\Big(\frac{m_h^2}{m_{KK}^2}\Big)$\\
$g_4=\frac{g_5}{\sqrt{2L}}$&$\tilde g_4=\frac{g_5}{\sqrt{2L}}\left(1-\co\Big(\frac{m_h^2}{m_{KK}^2}\Big)\right)$ & $\co\Big(\frac{m_h^2}{m_{KK}^2}\Big)$\\
$m_A^2=\frac{g_5^2}{2L} \frac{\mu^2}{\lambda}\left(1-\co\Big(\frac{m_h^2}{m_{KK}^2}\Big)\right)$&$\tilde m_A^2=\frac{g_5^2}{2L} \frac{\mu^2}{\lambda}\left(1-\co\Big(\frac{m_h^2}{m_{KK}^2}\Big)\right)$ & $\co\Big(\frac{m_h^2}{m_{KK}^2}\Big)$\\
\tabucline[1pt]{-}
\end{tabu}}
\caption{Comparison of the effective parameters in terms of the fundamental parameters of the 5D theory, where $\mu^2\equiv(1+\beta)m_{KK}^2\delta_{IR}$ and $\lambda=(1+\beta)^2\lambda_{IR}$. Here we explicitly show the presence of corrections of order of the expansion parameter,  $m_h^2/m_{KK}^2\sim10^{-3}$, and we neglect effects $\co\big(e^{-2\beta kL}\big)$;   the latter ones are extremely small for $\beta>0$ and $kL\sim35$.}
\label{comp_para}
\end{table}


\providecommand{\href}[2]{#2}\begingroup\raggedright\endgroup


\end{document}